\documentclass[preprint]{aastex6}
\newcommand{\wind}{\textit{Wind}}
\newcommand{\ace}{\textit{ACE}}
\newcommand{\ulysses}{\textit{Ulysses}}
\newcommand{\stereo}{\textit{STEREO}}
\newcommand{\soho}{\textit{SOHO${/}$}LASCO}
\newcommand{\genesis}{\textit{Genesis}}
\newcommand{\kms}{km s$^{-1}$}
\usepackage{amsmath}
\usepackage{graphicx}
\usepackage{array}
\usepackage{tabularx}

\shorttitle{Quantifying the propagation of seven fast CMEs}

\shortauthors{Zhao et al.}

\begin{document}

\title{Quantifying the Propagation of Fast Coronal Mass Ejections from the Sun to Interplanetary Space Combining Remote Sensing and Multi-Point in-situ Observations}

\author{Xiaowei Zhao\altaffilmark{1,2}, Ying D. Liu\altaffilmark{1,2}, Huidong Hu\altaffilmark{1}, and Rui Wang\altaffilmark{1}}

\altaffiltext{1}{State Key Laboratory of Space Weather, National Space Science Center, Chinese Academy of Sciences, Beijing 100190, China; \href{mailto:liuxying@swl.ac.cn}{liuxying@swl.ac.cn}}

\altaffiltext{2}{University of Chinese Academy of Sciences, Beijing 100049, China}

\begin{abstract}

In order to have a comprehensive view of the propagation and evolution of coronal mass ejections (CMEs) from the Sun to deep interplanetary space beyond 1 au, we carry out a kinematic analysis of 7 CMEs in solar cycle 23. The events are required to have coordinated coronagraph observations, interplanetary type II radio bursts, and multi-point in-situ measurements at the Earth and \ulysses. A graduated cylindrical shell model, an analytical model without free parameters and a magnetohydrodynamic model are used to derive CME kinematics near the Sun, to quantify the CME/shock propagation in the Sun-Earth space, and to connect in-situ signatures at the Earth and \ulysses, respectively. We find that each of the 7 CME-driven shocks experienced a major deceleration before reaching 1 au and thereafter propagated with a gradual deceleration from the Earth to larger distances. The resulting CME/shock propagation profile for each case is roughly consistent with all the data, which verifies the usefulness of the simple analytical model for CME/shock propagation in the heliosphere. The statistical analysis of CME kinematics indicates a tendency that the faster the CME, the larger the deceleration, and the shorter the deceleration time period within 1 au. For several of these events, the associated geomagnetic storms were mainly caused by the southward magnetic fields in the sheath region. In particular, the interaction between a CME-driven shock and a preceding ejecta significantly enhanced the preexisting southward magnetic fields and gave rise to a severe complex geomagnetic storm.

\end{abstract}

\keywords{Heliosphere --- Interplanetary shocks --- Solar wind --- Solar coronal mass ejections --- Solar radio emission}

\section{Introduction}

Coronal mass ejections are large expulsions of magnetized plasma and energy from the solar atmosphere. They are known as interplanetary CMEs (ICMEs) when moving through the heliosphere. The arrival of ICMEs with preceding shocks at the Earth can cause large geomagnetic storms \citep[e.g.,][]{1993Gosling, 1997Tsurutani}. How CMEs propagate in the heliosphere is important for space weather forecasting and research. Propagation of CMEs/shocks in the Sun-Earth space has been studied in recent years using wide-angle imaging observations from \stereo. However, their interplanetary transport beyond 1 au is still not well understood due to the lack of observations. Investigation of CME evolution beyond 1 au requires multi-point remote sensing and in-situ observations together with modeling techniques.

By combining Doppler scintillation, \textit{Solwind} coronagraph and \textit{Helios 1} plasma measurements, \citet{1985Woo} indicate that fast CME-driven shocks (with speeds exceeding 1000 \kms) may undergo a substantial deceleration near the Sun. \citet{2001GopalswamyL} propose that CMEs on average have a constant deceleration before 0.76 au and then comove with the ambient solar wind. It is difficult to fully understand CME propagation due to the large data gap between the Sun and the Earth. Based on LASCO and interplanetary scintillation images, \citet{2006Manoharan} suggests that fast CMEs are decelerated while slow ones are accelerated by the surrounding solar wind in the inner heliosphere, which is in agreement with previous findings \citep[e.g.,][]{1999Sheeley, 2000Gopalswamy}. With the use of the radio, in-situ, and white-light observations, \citet{2007Reiner} provide a kinematic analysis of 42 CMEs/shocks and suggest that a CME can cease its deceleration at anywhere from 0.27 au to about 1 au. A typical Sun-to-Earth propagation profile of fast CMEs has been found by \citet{2013Liu} based on stereoscopic wide-angle heliocentric imaging observations from \stereo. In their studies, CMEs first accelerate impulsively, then decelerate rapidly (out to 40--80 solar radii) due to interactions with the ambient solar wind, and finally move with a generally constant speed or gradual deceleration. For slow CMEs, \citet{2016Liu} point out that they are gradually accelerated to the ambient solar wind speed within 20--30 solar radii and thereafter comove with the ambient solar wind.

Interactions with other CMEs and solar wind structures can make the interplanetary propagation of CMEs/shocks more complicated. A CME can be deflected due to its interaction with coronal holes and other CMEs \citep[e.g.,][]{2001GopalswamyY, 2009Gopalswamy, 2012Lugaz, 2014LiuL}. \citet{2012Temmer} argue that a fast CME undergoes a strong deceleration when it catches up with a preceding CME. \citet{2014LiuL} suggest that an earlier CME erupting at an appropriate time plays a key role for the minor deceleration of a later CME by preconditioning the upstream solar wind. The preconditioning process is of importance for CME propagation, which has been verified by \citet{2015Cash} and \citet{2015Temmer}. \citet{2017Temmer} propose that the interplanetary medium needs 2--5 days to recover from the effects of preconditioning. In addition, high-speed solar wind streams can accelerate slow CMEs by compressing them from behind \citep{2015Kataoka, 2015Liu, 2016Liu}.

Ulysses, which was launched in 1990, provides an opportunity to investigate CME/shock propagation characteristics beyond 1 au. Based on coronagraph and \ulysses{} in-situ observations, \citet{1999Funsten} suggest that a CME can experience a deceleration when traveling from the Sun to \ulysses. By comparing the speeds of 11 ICMEs observed both at the Earth and \ulysses, \citet{2014Richardson} argues that fast CMEs with a higher speed than that of the ambient medium slowly decelerate beyond 1 au, which is consistent with the suggestion of \citet{2013Liu}. Combining coronagraph observations, interplanetary type II radio bursts, and in-situ measurements at the Earth and \ulysses, \citet{2008LiuL, 2017Liu} and \citet{2017Zhao} find that a fast CME/shock finished its major deceleration within 1 au and thereafter moved with a roughly constant speed. As far as we know, there are few statistical analyses of CME/shock propagation from the Sun to the Earth and then beyond 1 au using coordinated coronagraph observations, type II bursts, and in-situ measurements.

In this work, we provide a comprehensive study of the propagation and evolution of 7 CMEs from the Sun to deep interplanetary space beyond 1 au combining coordinated coronagraph observations, type II radio bursts, and in-situ solar wind measurements at the Earth and \ulysses{} with modeling techniques. We illustrate how to quantify the propagation of CMEs/shocks using the merged data sets, which is important for space weather forecasting. We also carry out a statistical kinematic analysis of the 7 CMEs in order to evaluate how fast CMEs decelerate in the heliosphere. The organization of this paper is as follows. We introduce the observations and methodology in Section \ref{obs}. Case studies are presented in Section \ref{eve}. Statistical analysis of CME kinematics is given in Section \ref{stat}. The results are summarized and discussed in Section \ref{sum}.

\section{Observations and methodology}\label{obs}

A primary purpose of this work is to examine how fast CMEs propagate from the Sun far into interplanetary space beyond 1 au. We use \soho{} coronagraph observations to investigate CME characteristics near the Sun. Radio type II burst observations covering the range from 20 kHz to 13.825 MHz from \wind/WAVES are analyzed in order to track the Sun-to-Earth propagation. Solar wind plasma and magnetic field data from \wind, \ace, \genesis{} \citep{2003Burnett} and \ulysses{} are used to identify corresponding ICMEs/shocks at different locations. Wind, \ace{} and \genesis{} are all at L1, and they complement each other when there are data gaps. We note that a few CMEs can be traced to the distance of the \textit{Voyager} spacecraft \citep[e.g.,][]{2001Wang, 2002Richardson, 2006Richardson, 2014LiuR}. Here we focus on the events using the \ulysses{} data. No \stereo{} wide-angle imaging observations are available for the \ulysses{} events, so we use long-duration interplanetary type II radio bursts instead. Type II burst observations are not imaging, but can be used to track interplanetary shocks (see below). Some events can also be covered by imaging observations from SMEI together with \ulysses{} data \citep[e.g.,][]{2005Reiner, 2006Tappin}, but the use of SMEI data is beyond the scope of our paper. The methods applied in this paper are introduced as follows.

\subsection{Graduated Cylindrical Shell Model}\label{gcs}

The graduated cylindrical shell (GCS) model is a forward modeling technique proposed by \citet{2006Thernisien, 2009Thernisien} and can reproduce CMEs with a croissant-like morphology based on coronagraph observations \citep[e.g.,][]{2010Liu, 2017Liu, 2019Liu, 2013Cheng, 2015Mishra, 2017Hu}. This model has six free parameters: the longitude and latitude of the propagation direction, the tilt angle, half angle and aspect ratio of the flux rope, and the heliocentric distance of the CME leading front. In this work, we keep the tilt angle, aspect ratio and the half angle parameters (roughly) unchanged and slightly adjust the longitude and latitude of the flux rope. The tilt angle is selected to agree with the orientation of the neutral line in the source active region (AR). By applying this model to \soho{} coronagraph observations, we can estimate CME propagation direction and initial speed near the Sun.

\subsection{Frequency Drift of Type II Radio Burst}\label{freq}

Type II bursts are radio emissions by shock-accelerated electrons near the local plasma frequency and/or its harmonics \citep[e.g.,][]{1985Nelson, 1987Cane}. As remote signatures of CME-driven shocks, they drift to lower frequencies as the shocks propagate away from the Sun. The frequency associated with the type II radio burst can be converted to the CME/shock propagation distance based on a suitable solar wind density model \citep[e.g.,][]{1976Chertok, 2007Reiner, 2008LiuL, 2013Liu, 2017Liu, 2015Cremades, 2016Hu, 2017Zhao}. Using an electron density model \citep[][referred to as the Leblanc density model hereafter]{1998Leblanc}, \citet{2013Liu} obtain the radial distances of CME-driven shocks from the associated type II bursts, which are consistent with the results from wide-angle stereoscopic imaging observations. In this work, we employ the same density model to convert the frequency of the associated type II burst to propagation distance by adjusting the nominal density at 1 au.

\subsection{Analytical Model of CME/Shock Propagation}\label{prop}

We use a simple analytical model with no free parameters proposed by \citet{2017Liu} to quantify the propagation of CME-driven shocks within 1 au. The shock is assumed to start with an initial speed ${v}_{0}$ at a time ${t}_{0}$ and a distance ${r}_{0}$ from the Sun, propagate with a constant deceleration ${a}$ for a time period ${t}_{a}$ before reaching 1 au, and afterwards move with a constant speed ${v}_{s}$. The initial speed ${v}_{0}$ can be calculated from a linear fit of the heliocentric distances given by the GCS model. For simplicity, the distance ${r}_{0}$ is set to 0 and the time ${t}_{0}$ is set to be the mid-time between the flare start and the flare maximum. The shock speed ${v}_{s}$ and the transit time ${t}_{s}$ are obtained from the in-situ measurements at the Earth. Combining these known parameters, we can derive key parameters of CME/shock propagation inside 1 au, including the time for deceleration ${t}_{a}=2\left({d}_{s}-{v}_{s}{t}_{s}\right)/\left({v}_{0}-{v}_{s}\right)$, where ${d}_{s}$ is the distance of \wind/\ace{} from the Sun, the deceleration ${a}=\left({v}_{s}-{v}_{0}\right)/{t}_{a}$, and the deceleration cessation distance ${r}_{a}=\left({v}_{0}+{v}_{s}\right){t}_{a}/2$. We can use the frequency drift of the associated type II burst to verify the height-time profile derived from the analytical model.

\subsection{Magnetohydrodynamic Model}\label{mhd}

We use a 1D magnetohydrodynamic (MHD) model developed by \citet{2000Wang} to investigate the propagation of solar wind disturbances beyond 1 au. This MHD model has been used in previous studies with an assumption of spherical symmetry since solar wind measurements are one-dimensional \citep[e.g.,][]{2001Wang, 2002Richardson, 2006Liu, 2008LiuL, 2017Liu, 2007Du, 2017Zhao}. We apply the MHD model to hourly averages of the solar wind data from the near-Earth spacecraft (\wind, \ace{} and \genesis) in order to check whether the measurements at the Earth and \ulysses{} are consistent with the propagation of the same ICME/shock. The MHD model results can also be used to evaluate the analytical model (see Section \ref{prop}) beyond 1 au.

\section{Event Analysis}\label{eve}

We perform a survey of CMEs/ICMEs, which we require to have coronagraph observations from \soho, in-situ solar wind measurements at the Earth and \ulysses, and long duration interplanetary type II radio bursts from \wind. The survey yields 7 such events. Table \ref{tab} gives the details of 7 CMEs, including the eruption date, associated AR and flare intensity, CME propagation direction near the Sun, shock arrival time at the Earth and \ulysses, the latitude and longitude of \ulysses{} relative to the Earth, and the distance of \ulysses{} from the Sun. The CME propagation directions are generally consistent with the source locations on the Sun, but we do see deviations from the source regions. Here we describe detailed studies of the first 4 CMEs, since the remaining 3 CMEs have been studied in previous works \citep{2008LiuL, 2017Liu, 2017Zhao}.

\subsection{The 1997 November 4 CME}

\begin{figure}
\epsscale{1.1} \plottwo{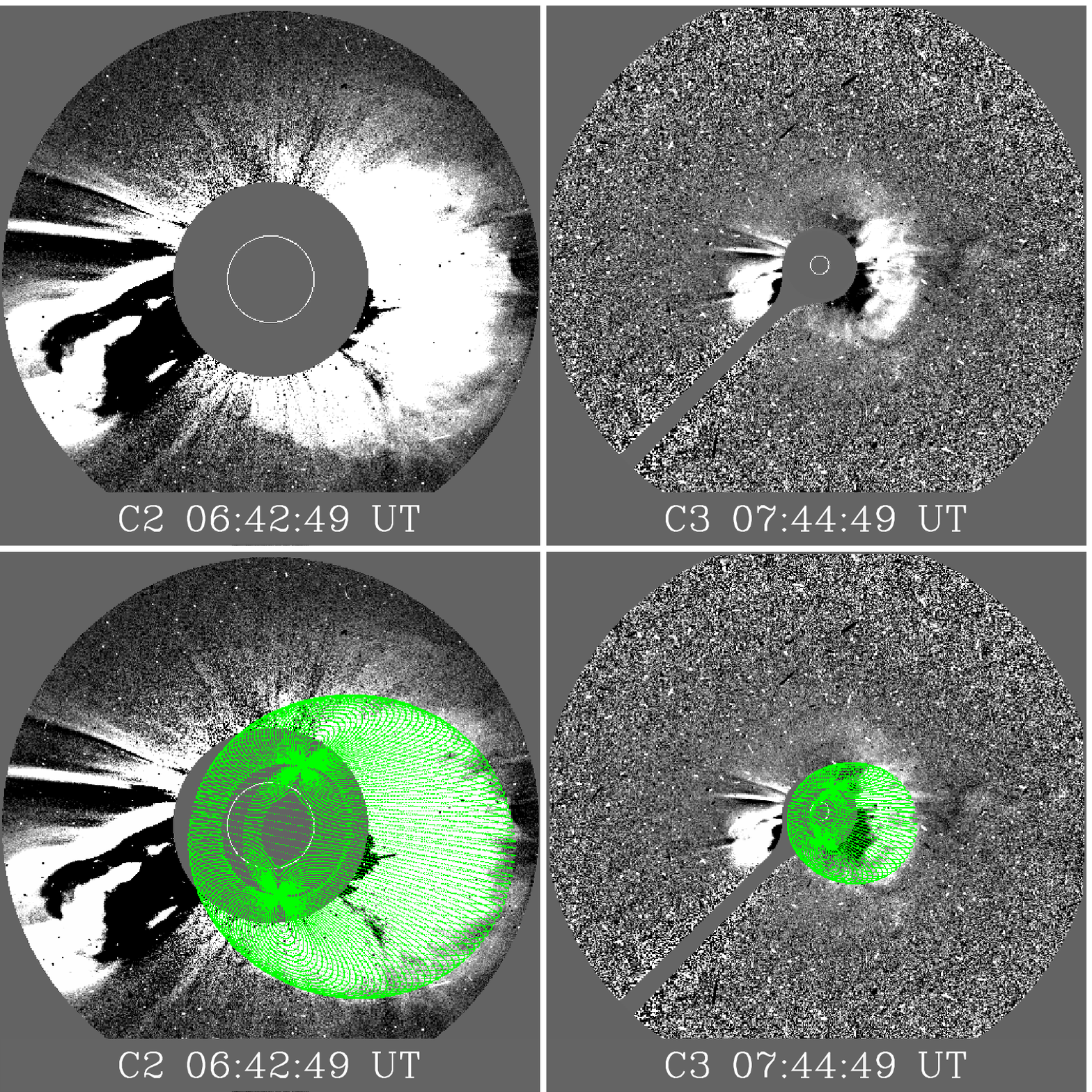}{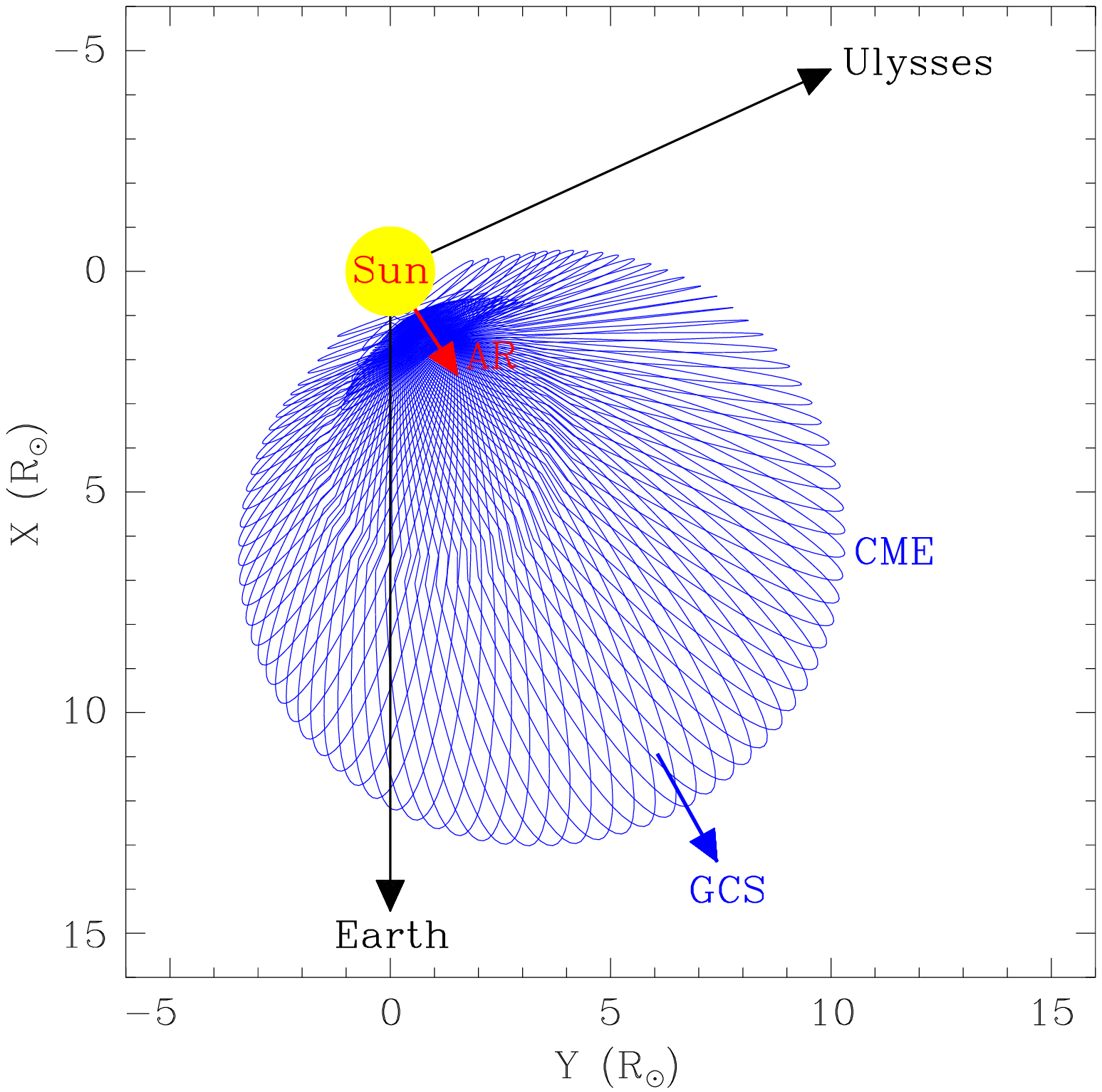}
\caption{\label{f1}Left: Coronagraph running-difference images from LASCO and corresponding GCS modeling (green grids) for the 1997 November 4 CME. Right: Projection of the modeled CME at 07:44:49 UT onto the ecliptic plane. The red arrow indicates the solar source location (S14{\degr}W33{\degr}). The blue arrow shows the CME propagation direction from the GCS model (S04{\degr}W29{\degr}). Ulysses was 2{\degr} south and 114.6{\degr} west of the Earth when the CME erupted from the Sun (see Table \ref{tab}). }
\end{figure}

On 1997 November 4, a CME launched from NOAA AR 08100 (S14{\degr}W33{\degr}) and was associated with an X2.1 flare peaking at 05:58 UT. Figure \ref{f1} (left) shows \soho{} coronagraph observations (C2 and C3) and GCS modeling. The CME propagating to the west is the event of interest. The faint edge around the CME front is the CME-driven shock. The flux-rope morphologies of the CME at different times are reproduced by applying the GCS model, which are visually consistent with the white light images. The application of this model to single spacecraft observations may give rise to large uncertainties in the parameters. Here we use it to get rough estimates of CME propagation direction and speed near the Sun. The CME was propagating along 4{\degr} south and 29{\degr} west of the Earth (see Table \ref{tab}), which is similar to its solar source location (S14{\degr}W33{\degr}). A linear fit of the CME leading front distances obtained from the GCS method gives an average speed of about 1200 \kms{} near the Sun. It is used as an approximation of the shock speed, since the shock is very close to the CME leading front near the Sun. Figure \ref{f1} (right) displays the projection of the modeled CME onto the ecliptic plane and the positions of the Earth and \ulysses{} when the CME erupted from the Sun. The CME should be observed at both the Earth and \ulysses{} given its propagation direction and width.

\begin{figure}
\epsscale{0.8} \plotone{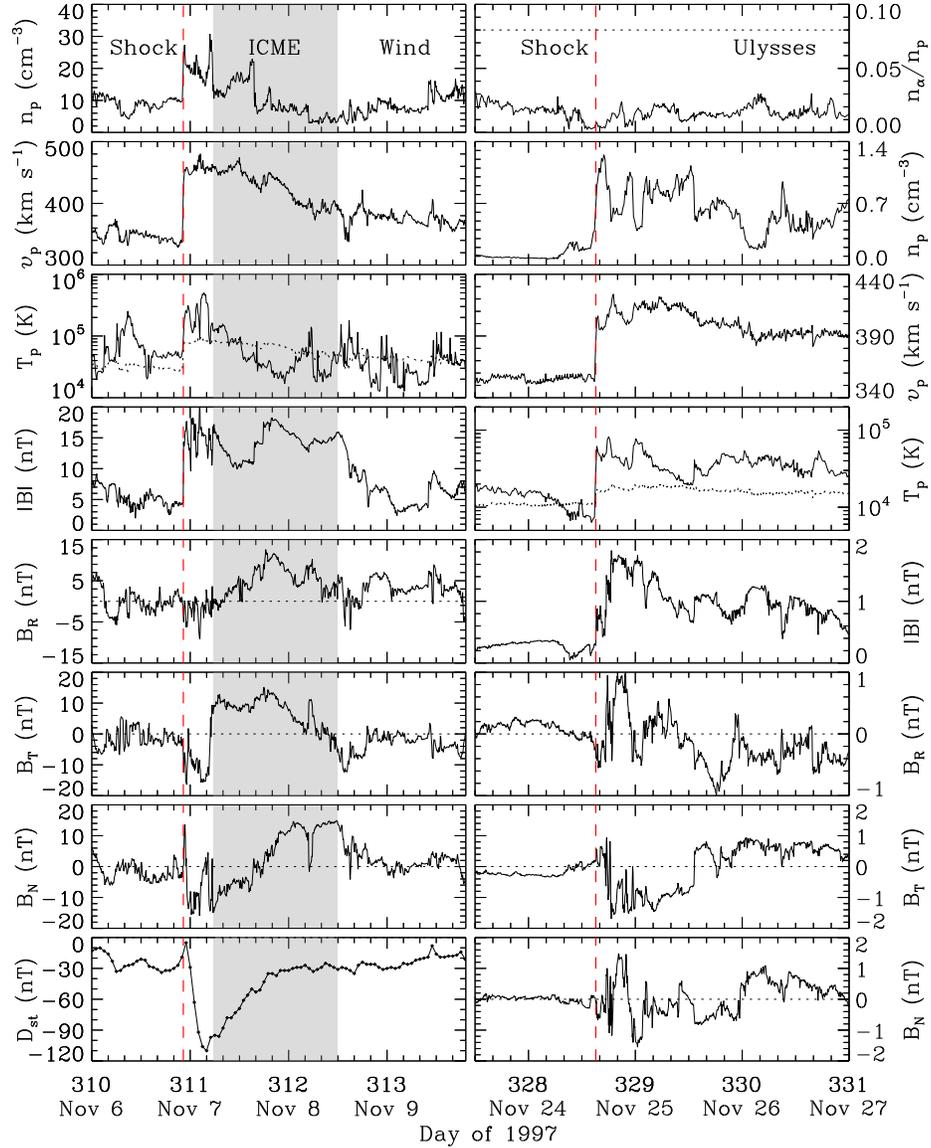}
\caption{\label{f2}Left: solar wind measurements of the 1997 November 4 CME at \wind{}. From top to bottom, the panels show proton density, bulk speed, proton temperature, magnetic field strength and components, and \textit{D}$_\mathrm{st}$ index, respectively. The dotted curve in the third panel denotes the expected proton temperature obtained from the observed solar wind speed \citep{1987Lopez}. The shaded region marks the interval of the identified ICME. Right: similar to left, but for solar wind measurements at \ulysses{}. The first panel gives the density ratio between alphas and protons, and the horizontal dotted line marks the 0.08 level, which can be used as a threshold for ICMEs \citep{2004Richardson, 2005Liu, 2006LiuR}. The dotted curve in the fourth panel is the expected proton temperature taking into account the temperature-distance gradient \citep[\textit{R}$^{-0.7}$;][]{1994Gazis}. The two red vertical dashed lines indicate the arrival times of the shock at \wind{} and \ulysses{}, respectively. }
\end{figure}

Figure \ref{f2} presents \wind{} (left) and \ulysses{} (right) in-situ measurements of the 1997 November 4 CME. The magnetic field data are in \textbf{RTN} coordinates, where \textbf{R} points from the Sun to the spacecraft, \textbf{T} lies in the solar equatorial plane and points towards the planet motion direction, and \textbf{N} completes the right-handed triad. We see a preceding shock passed \wind{} around 22:19 UT on November 6 with a speed of about 560 \kms. The shaded region shows the ICME interval determined from the smooth rotation of the magnetic field components and the depressed proton temperature compared with the expected temperature derived from a temperature--speed relationship \citep{1987Lopez}. The CME/shock transit time from the Sun to the Earth is about 64.4 hr, which gives an average speed of about 640 \kms. An intense geomagnetic storm with a minimum \textit{D}$_\mathrm{st}$ index of ${-}$110 nT occurred with the shock arrival at the Earth. The geomagnetic storm is mainly caused by the enhanced southward magnetic field (negative \textbf{N} components) first in the sheath region and then in the leading part of the ICME.

A shock was observed at \ulysses{} at about 15:07 UT on November 24 with a speed of about 410 \kms{} (see Figure \ref{f2}). Ulysses was 0.7{\degr} south and 94.8{\degr} west of the Earth and at a distance of 5.34 au from the Sun when it observed the shock (Table \ref{tab} gives the direction and distance of \ulysses{} at the time of the eruption). There are no obvious ICME signatures, such as enhanced helium abundance, depressed proton temperature, or smooth field rotations. The shock transit time from the Earth to \ulysses{} ($\sim$17.7 days) implies an average speed of about 430 \kms. The transit speed (430 \kms) agrees well with the \ulysses{} in-situ measurements, which indicates that the shock observed by \ulysses{} is probably the same shock at the Earth. Comparing the shock speeds near the Sun (1200 \kms), at the Earth (560 \kms), and at \ulysses{} (410 \kms) suggests that the major deceleration of the CME/shock occurred inside 1 au and then the CME/shock propagated with a gradual deceleration beyond 1 au. This is consistent with the findings of \citet{2013Liu, 2016Liu, 2017Liu}.

\begin{figure}
\epsscale{0.8} \plotone{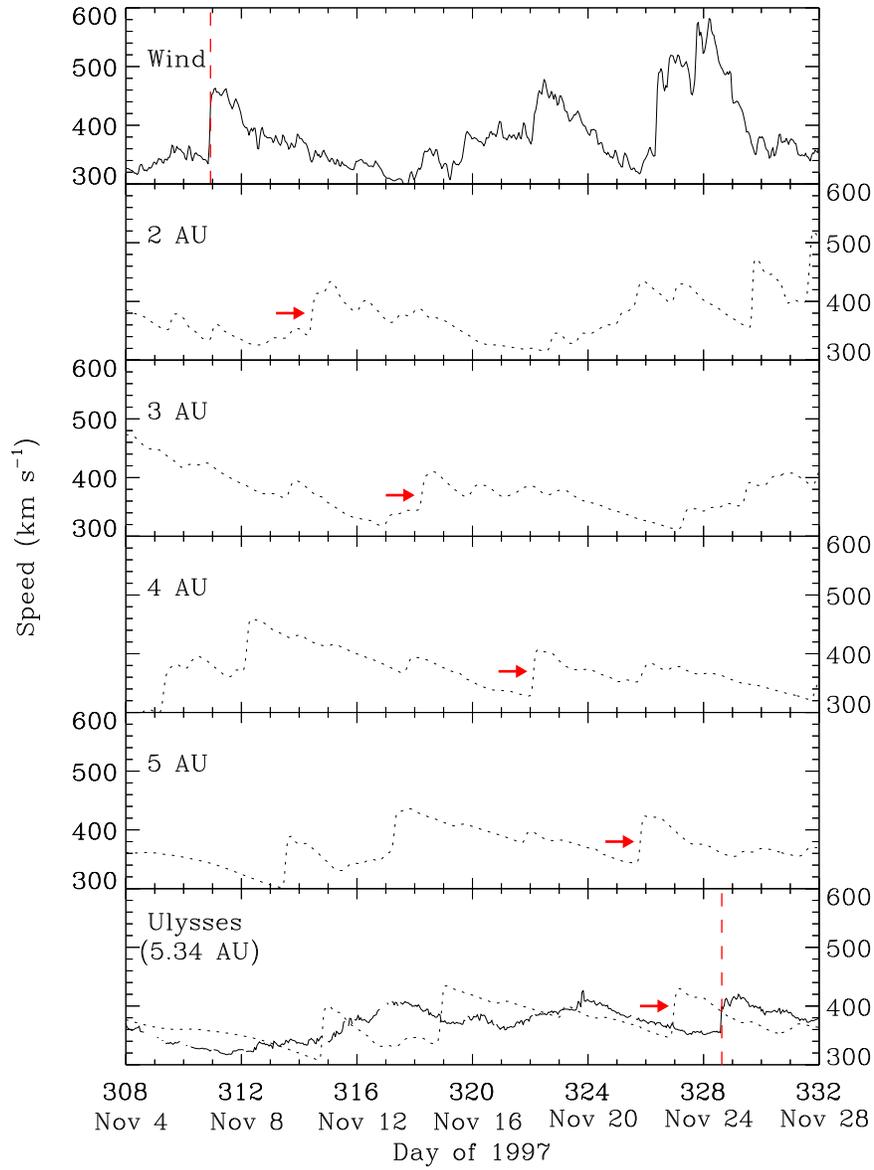}
\caption{\label{f3}Evolution of solar wind speeds from \wind{} to \ulysses{} via the 1D MHD model. The solid curves give the solar wind speeds observed at \wind{} and \ulysses{}, and the dotted curves represent the predicted speeds at specific distances. The red arrows indicate the progression of the modeled shock. The two red vertical dashed lines mark the observed shock arrival times at \wind{} and \ulysses{}, respectively. The same scales are used for all of the panels.}
\end{figure}

We propagate the solar wind disturbances at \wind{} outward with the MHD model (see Section \ref{mhd}), in order to connect with the in-situ measurements at \ulysses. Figure \ref{f3} displays the model output at given distances and the observed solar wind speeds at \wind{} and \ulysses, respectively. The shock persists during transit from the Earth to \ulysses. The predicted shock arrival time at \ulysses{} is about 00:00 UT on November 23, which is about 39 hr earlier than observed. The time difference (39 hr) is probably caused by the large longitudinal separation between the Earth and \ulysses. However, it is much smaller than the transit time from the Earth to \ulysses{} ($\sim$425 hr).

We note that another fast CME erupted from the same active region two days later. The November 6 CME missed the Earth, but the possibility of its shock arrival at \ulysses{} cannot be completely excluded, since its solar source location (S18{\degr}W63{\degr}) is closer to the direction of \ulysses{} compared with that of the November 4 CME (S14{\degr}W33{\degr}). However, the MHD model results can still be used to compare with the height-time profile of the November 4 CME beyond 1 au derived from the analytical model.

\begin{figure}
\epsscale{1.0} \plotone{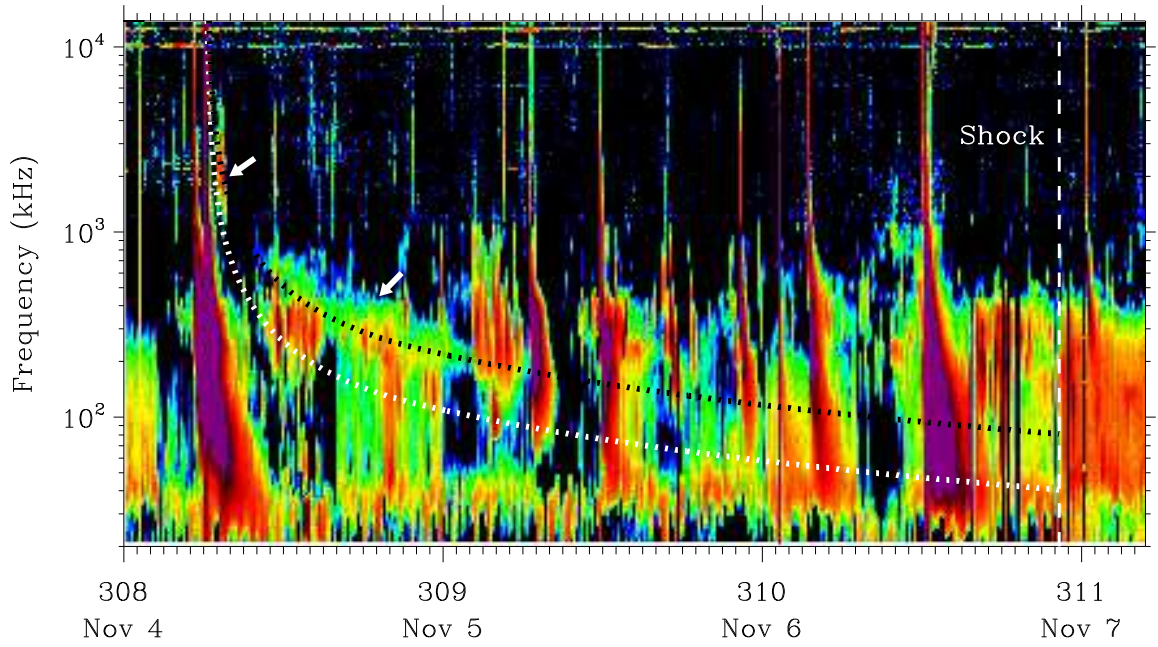}
\caption{\label{f4} Dynamic spectrum associated with the 1997 November 4 CME from \wind{}. The vertical dashed line indicates the shock arrival time at the Earth. The white arrows indicate the observed type II radio bands. The black dotted curve is obtained from the simple analytical model, which simulates the frequency drift of the observed type II radio burst at the harmonic plasma frequency. The white dotted curve is derived by dividing the harmonic curve by a factor of 2.  }
\end{figure}

A type II radio burst provides an opportunity to track a CME-driven shock from the Sun to the Earth and fill the gap of direct solar wind measurements \citep[e.g.,][]{2007Reiner, 2008LiuL, 2017Liu}. Figure \ref{f4} gives the radio dynamic spectrum associated with the 1997 November 4 CME. Note that type II bands can be noisy and difficult to identify (for example, due to the contamination by type III bursts). In the present case, the type II burst likely commenced at about 06:45 UT on November 4 and the frequency drifted downward possibly at both the fundamental and harmonic plasma frequencies until 07:38 UT on November 4. Then diffuse type II radio bands reappeared at about 09:36 UT on November 4 and split into two bands. After about 02:05 UT on November 5, there are a series of type III bursts, and no clear type II signatures are observed before the shock arrived at \wind. \citet{1999Dulk} suggest that the CME, the type II burst, and the shock near the Earth associated with this event are closely interrelated, which further confirms the connection between the CME near the Sun and the ICME at 1 au.

We apply the analytical model (see Section \ref{prop}) to quantify the shock propagation within 1 au. The shock speeds near the Sun (1200 \kms) and at the Earth (560 \kms) and the transit time (64.4 hr) are used as input to the model, which gives a deceleration ${a}=-11.0$ m s$^{-2}$, a deceleration time ${t}_{a}=16.2$ hr, and a deceleration cessation distance ${r}_{a}=0.34$ au. Then we convert the distances from the analytical model to frequencies employing the Leblanc density model \citep{1998Leblanc}. For the use of the Leblanc density model, we direct readers to \citet{2017Liu} and \citet{2017Zhao} who perform similar analyses. We adjust ${n}_{0}$ (the nominal 1-au electron density in the Leblanc model) in order to match the frequency drift of the harmonic type II bands. A value of ${n}_{0}=20$ cm$^{-3}$ yields a frequency-time profile that is consistent with the observed radio spectrum (see Figure \ref{f4}). This result is different from the study of \citet{1999Dulk} assuming that the type II emission was at the fundamental plasma frequency with a normalized density at 1 au of 12.6 cm$^{-3}$. However, the resulting CME kinematics are actually independent of the density model used in the present work.

\begin{figure}
\epsscale{1.0} \plotone{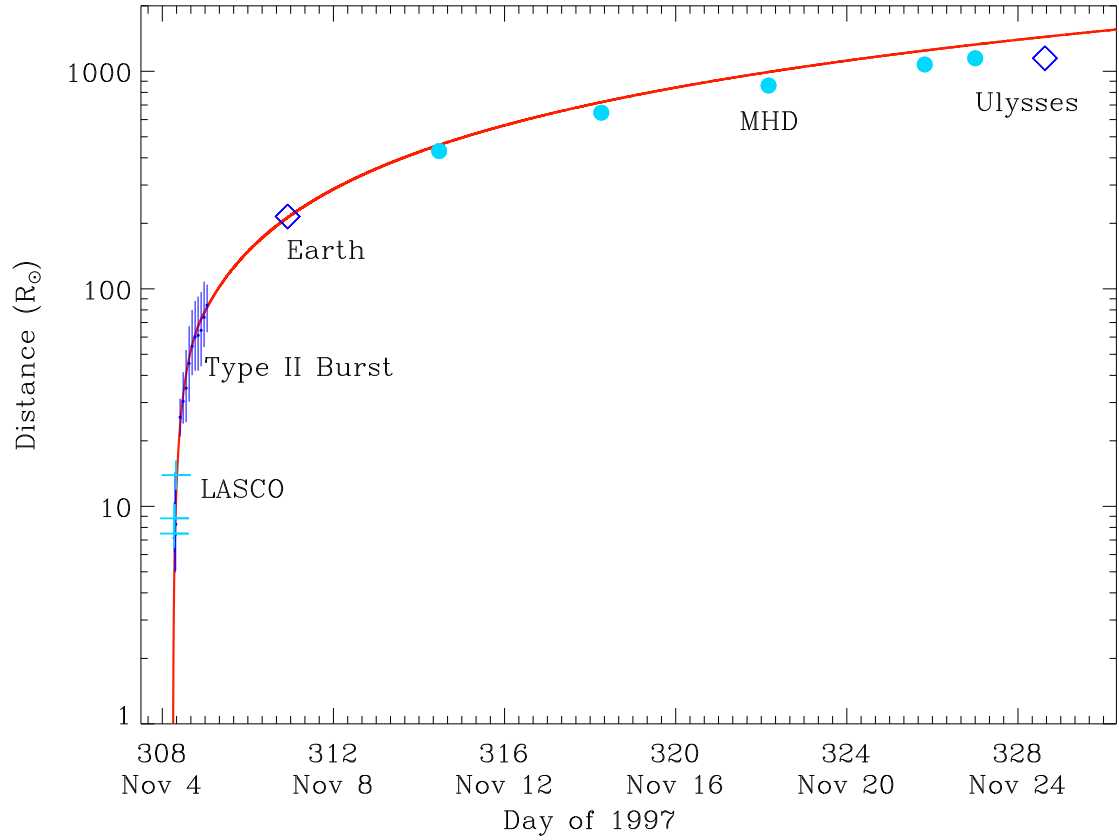}
\caption{\label{f5}Overall propagation profile of the 1997 November 4 CME-driven shock from the Sun to deep interplanetary space obtained from the analytical model (red curve). Crosses are distances of the CME leading front from the GCS model. Small dots with error bars denote the distances determined from the observed type II burst based on the Leblanc density model \citep{1998Leblanc}. The error bars correspond to the width of the type II band. Diamonds show the distances of the Earth and \ulysses{} at shock arrival, respectively. Large blue dots indicate the arrival times of the shock at 2, 3, 4, 5, and 5.34 au (\ulysses{}) predicted by the MHD model. }
\end{figure}

Plotted in Figure \ref{f5} are distances obtained from the GCS modeling of coronagraph observations, the observed type II bursts, in-situ measurements of the shock arrival at \wind{} and \ulysses, and the MHD model output. We extend the propagation profile from the analytical model to \ulysses{} in order to  show whether it accords with the distances beyond 1 au. The height-time curve, in general, agrees with all the distances from the Sun to \ulysses, but slightly deviates from the MHD model output at larger distances and the distance of \ulysses. Again, this probably results from the large longitudinal separation between the Earth and \ulysses{} in combination with the assumption of a constant speed beyond 1 au in the analytical model. Overall, the height-time profile is generally consistent with the rapid and gradual deceleration phases for fast CMEs found by \citet{2013Liu}. The deceleration cessation distance (0.34 au) is smaller than the result (0.56 au) obtained by \citet{2007Reiner} for the same CME, and it falls in the range of 0.2--0.4 au given by \citet{2013Liu} for three fast CMEs.

\subsection{The 2000 June 6 CME}

\begin{figure}
\epsscale{1.1} \plottwo{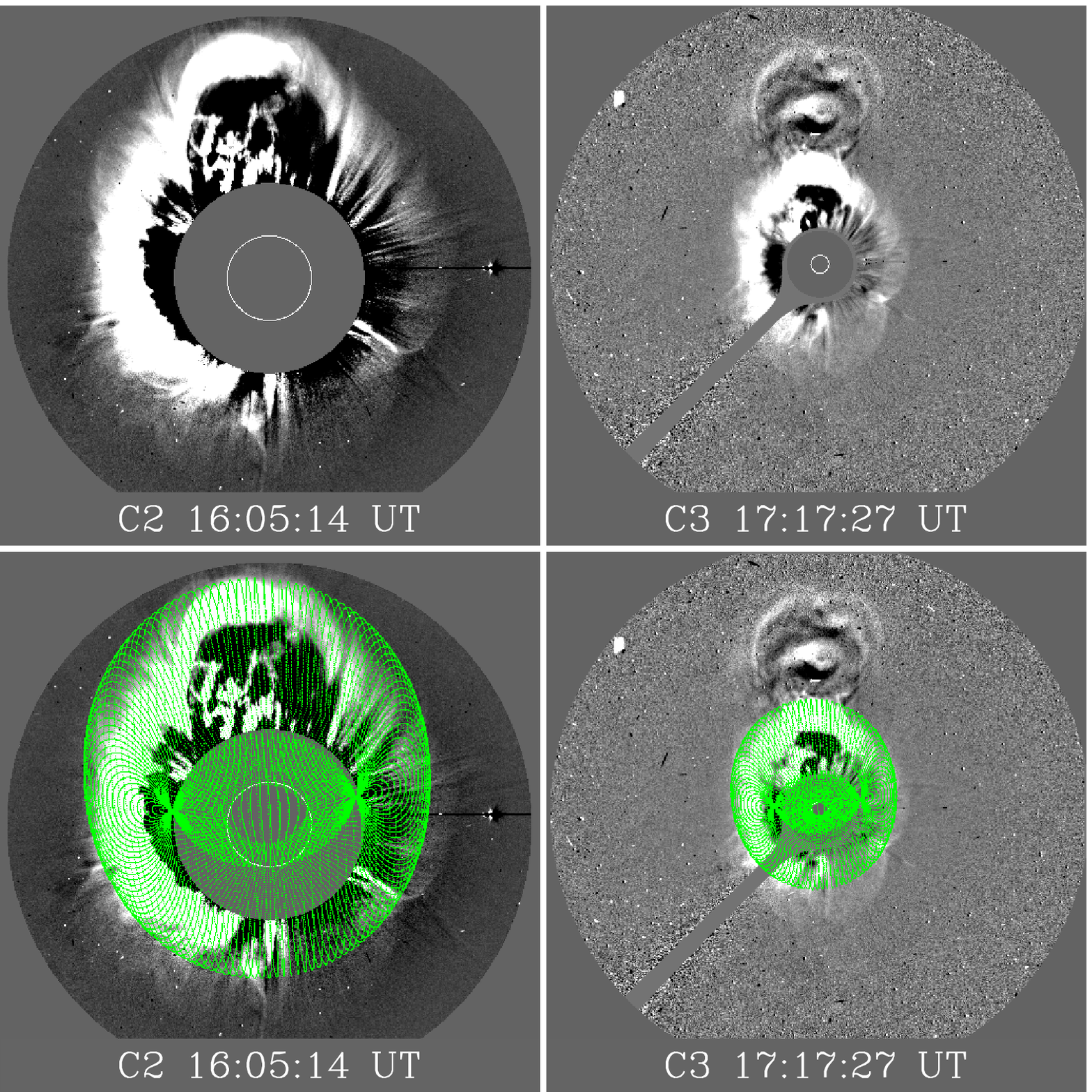}{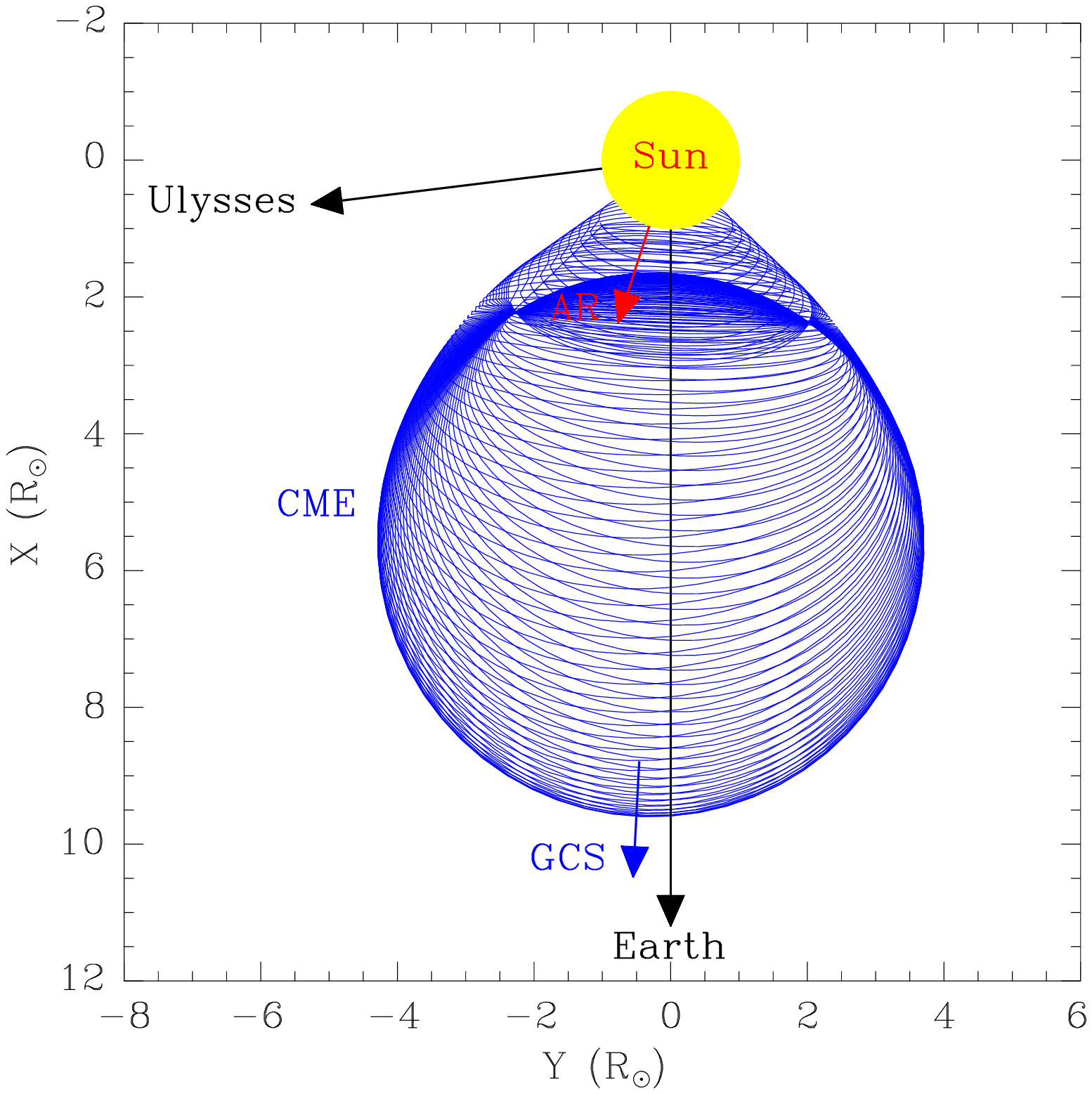}
\caption{\label{f6}Similar to Figure \ref{f1}, but for the 2000 June 6 CME. The modeled CME at 16:05:14 UT is projected onto the ecliptic plane. Ulysses was at 57.4{\degr} south and 83.0{\degr} east of the Earth when the CME launched from the Sun (see Table \ref{tab}).}
\end{figure}

A CME erupted from NOAA AR 09026 (N20{\degr}E18{\degr}) on 2000 June 6 and was accompanied by an X2.3 flare that peaked at 15:25 UT. Figure \ref{f6} (left) presents coronagraph observations and modeling of the 2000 June 6 CME. Again, we see a shock signature ahead of the CME front. There is a good consistency between the modeled flux-rope outlines and the observed images as in the previous event. The CME propagation direction near the Sun is N11{\degr}E03{\degr}, which agrees with its solar source location (N20{\degr}E18{\degr}). An average speed (1800 \kms), derived from a linear fit of the GCS distances, is used as a proxy for the shock speed near the Sun. Figure \ref{f6} (right) gives the projection of the modeled flux-rope morphology onto the ecliptic plane and the relative positions of the Earth and \ulysses. The CME may impact the Earth head on and encounter \ulysses{} with its flank, since its direction is much closer to the direction of the Earth.

\begin{figure}
\epsscale{0.9} \plotone{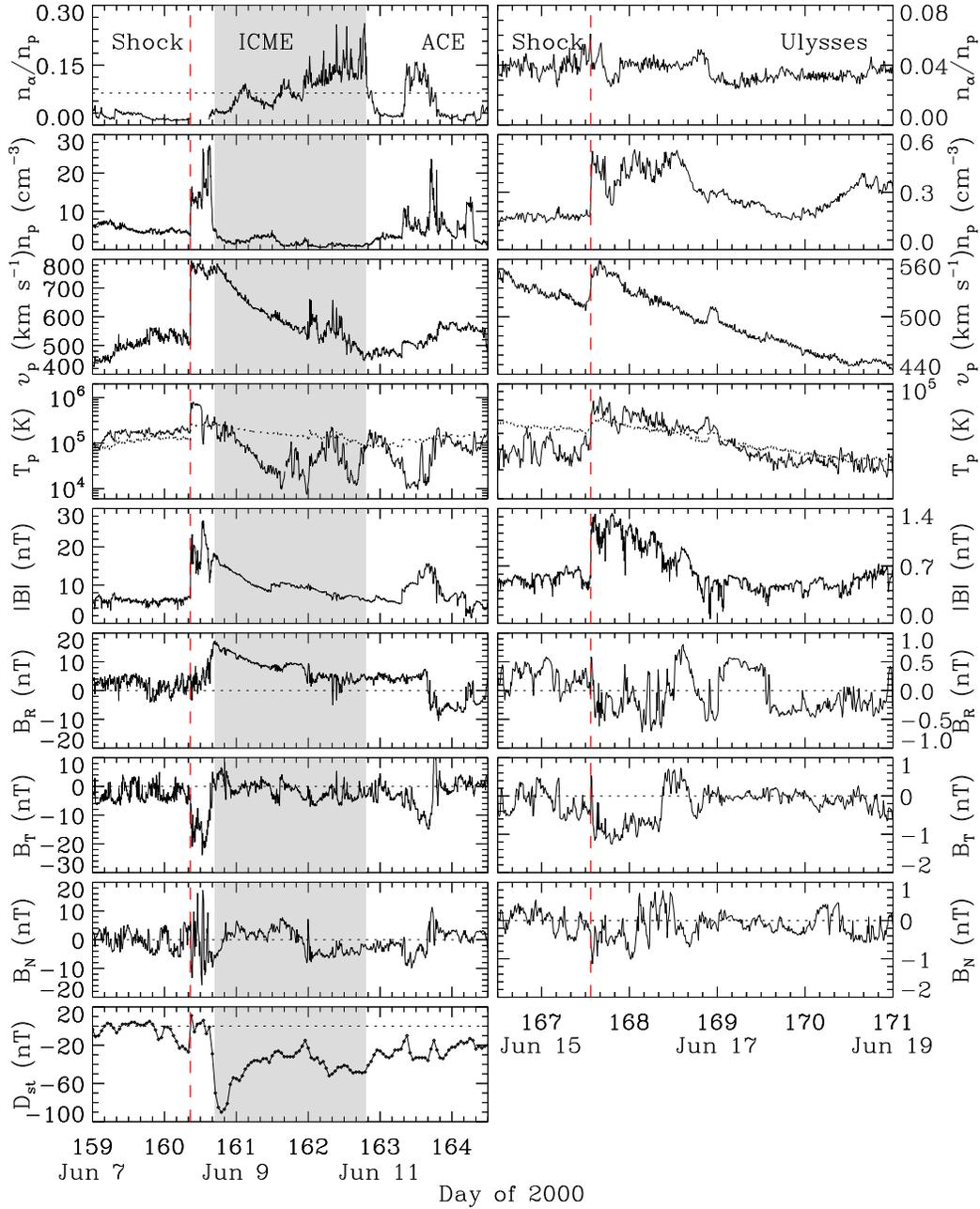}
\caption{\label{f7}Similar to Figure \ref{f2}, but for the in-situ measurements of the 2000 June 6 CME. Data from \ace{} have been used due to the deficiency of the solar wind measurements at \wind{}. The first panel on the left shows the density ratio between alphas and protons with the dotted line marking the 0.08 level. }
\end{figure}

Figure \ref{f7} shows the in-situ solar wind measurements at \ace{} (left) and \ulysses{} (right) for the 2000 June 6 CME. At about 08:38 UT on June 8, a forward shock arrived at \ace{} with a speed of about 890 \kms. The shaded region shows the identified ICME with an interval from 16:48 UT on June 8 to 19:12 UT on June 10, which is slightly longer than that of \citet{2009Rodriguez}. The ICME interval is mainly determined from the enhanced alpha-to-proton density ratio, low proton density, and depressed proton temperature. The time duration from the eruption to the shock arrival at \ace{} ($\sim$41.4 hr) gives an average transit speed of about 1020 \kms. A geomagnetic storm occurred with a minimum \textit{D}$_\mathrm{st}$ index of ${-}$90 nT, which is mainly caused by the southward magnetic field component in the sheath region and the leading part of the ejecta. The trailing part of the ejecta also has southward fields, which gave rise to another dip in the \textit{D}$_\mathrm{st}$ index although weak. Therefore, this is a multi-step geomagnetic storm.

Ulysses observed a shock at about 13:26 UT on June 15 at a distance of 3.34 au from the Sun. It was 90.9{\degr} east and 59.5{\degr} south of the Earth when the shock arrived there. There are no clear ICME signatures behind the shock. The time interval between the shock arrival at \ace{} and \ulysses{} is about 7.2 days, which gives an average transit speed of about 560 \kms. The shock at \ulysses{} is probably the same shock at \ace, since the transit speed (560 \kms) is similar to the shock speed at \ulysses{} (570 \kms). Comparison between the shock speeds near the Sun (1800 \kms), at the Earth (890 \kms), and at \ulysses{} (570 \kms) indicates that the shock finished its major deceleration inside 1 au and thereafter slowly decelerated beyond 1 au.

\begin{figure}
\epsscale{0.8} \plotone{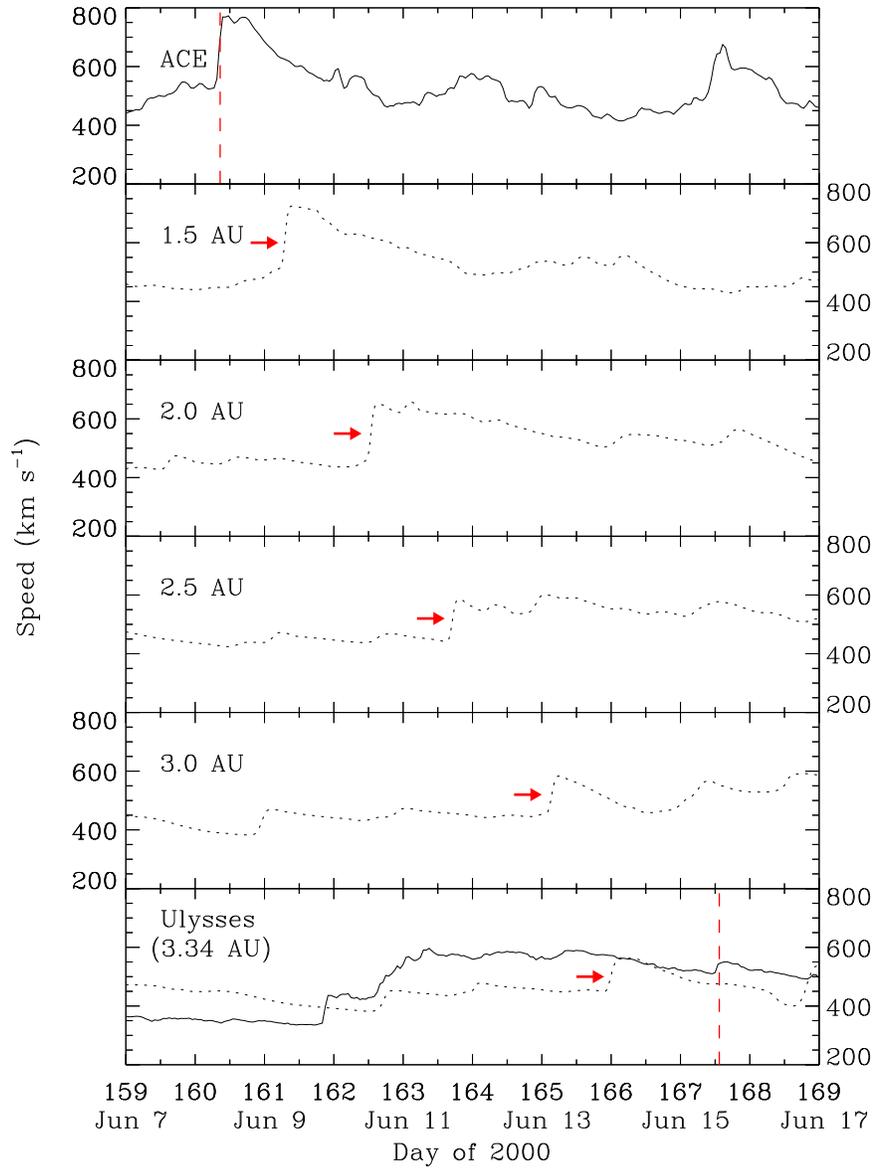}
\caption{\label{f8}Similar to Figure \ref{f3}, but for the solar wind speed evolution of the 2000 June 6 CME from \ace{} to \ulysses{} via the 1D MHD model.}
\end{figure}

Figure \ref{f8} displays solar wind speeds output by the MHD model and the observed ones at \ace{} and \ulysses{}, respectively. The shock is persistent during the transit from the Earth to \ulysses. The predicted arrival time of the shock at \ulysses{} is around 00:28 UT on June 14, which is about 37 hr earlier than observed. The time difference (37 hr) is much smaller than the transit time from the Earth to \ulysses{} ($\sim$173 hr), and is acceptable given the relative positions of the Earth and \ulysses. Again, the time difference probably results from the large longitudinal and latitudinal separations between the Earth and \ulysses. Note that the observed speed around the shock at \ulysses{} is higher than the model speed, which is reasonable since \ulysses{} was at a high latitude (S58.4{\degr}).

\begin{figure}
\epsscale{1.0} \plotone{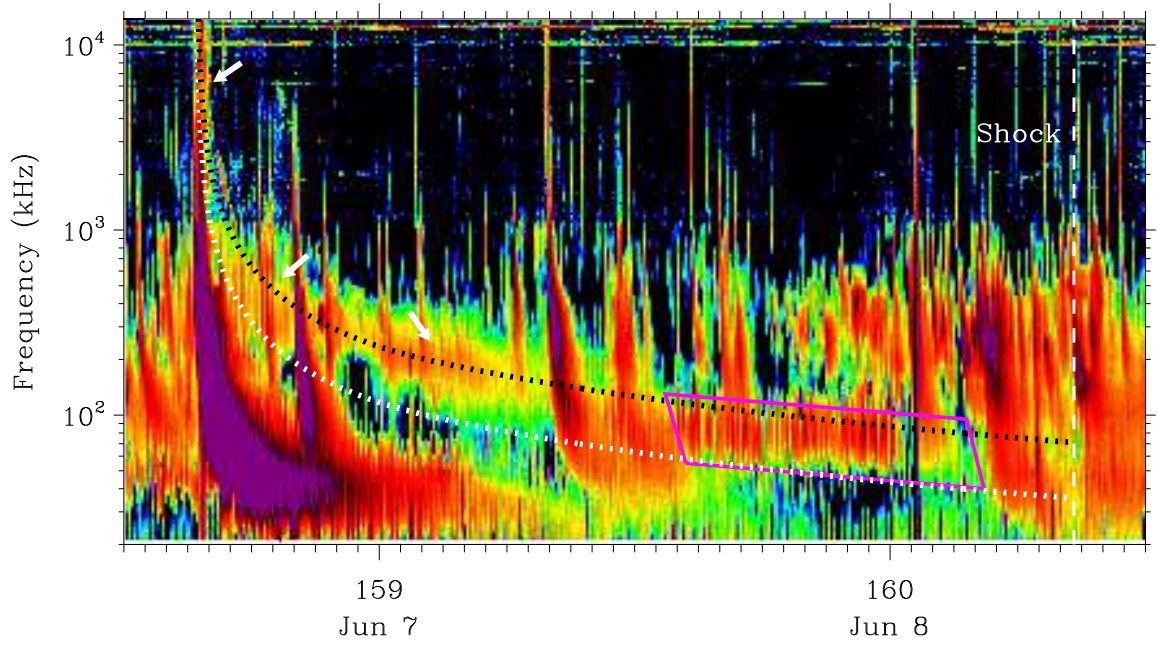}
\caption{\label{f9}Similar to Figure \ref{f4}, but for the dynamic spectrum associated with the 2000 June 6 CME. The white arrows indicate the observed type II radio bands. The pink quadrilateral shows a special region, covering frequencies from about 40 {\rm{kHz}} to 130 {\rm{kHz}} and times from about 13:26 UT on June 7 to 04:25 UT on June 8. }
\end{figure}

The radio dynamic spectrum associated with the 2000 June 6 CME is given in Figure \ref{f9}. Diffuse type II bands were observed at both the fundamental and harmonic plasma frequencies from about 15:36 UT to 22:12 UT on June 6 and thereafter mainly seen at the harmonic plasma frequency until 07:55 UT on June 7. A region indicated by the pink quadrilateral in Figure \ref{f9} might be part of the type II burst at the harmonic plasma frequency, but its identification is not clear due to contamination by type III emissions. The type II radio burst associated with the 2000 June 6 CME has been also studied by \citet{2007Reiner}, which confirms the link between the CME near the Sun and the ICME at 1 au.

The shock propagation is characterized by the analytical model (see Section \ref{prop}). With the known parameters, including the shock speeds near the Sun (1800 \kms) and at the Earth (890 \kms) and the transit time (41.4 hr), the model yields a deceleration ${a}=-22.1$ m s$^{-2}$, a time for deceleration ${t}_{a}=11.4$ hr, and a deceleration cessation distance ${r}_{a}=0.37$ au. Then the distances from the analytical model are converted to frequencies using the Leblanc density model. A scale of ${n}_{0}=16$ cm$^{-3}$ in the density model gives a frequency-time profile that accords with the observed type II emissions at the harmonic plasma frequency. The obtained profile is roughly consistent with the special region but slightly higher than the average frequency in this region (see Figure \ref{f9}).

\begin{figure}
\epsscale{1.0} \plotone{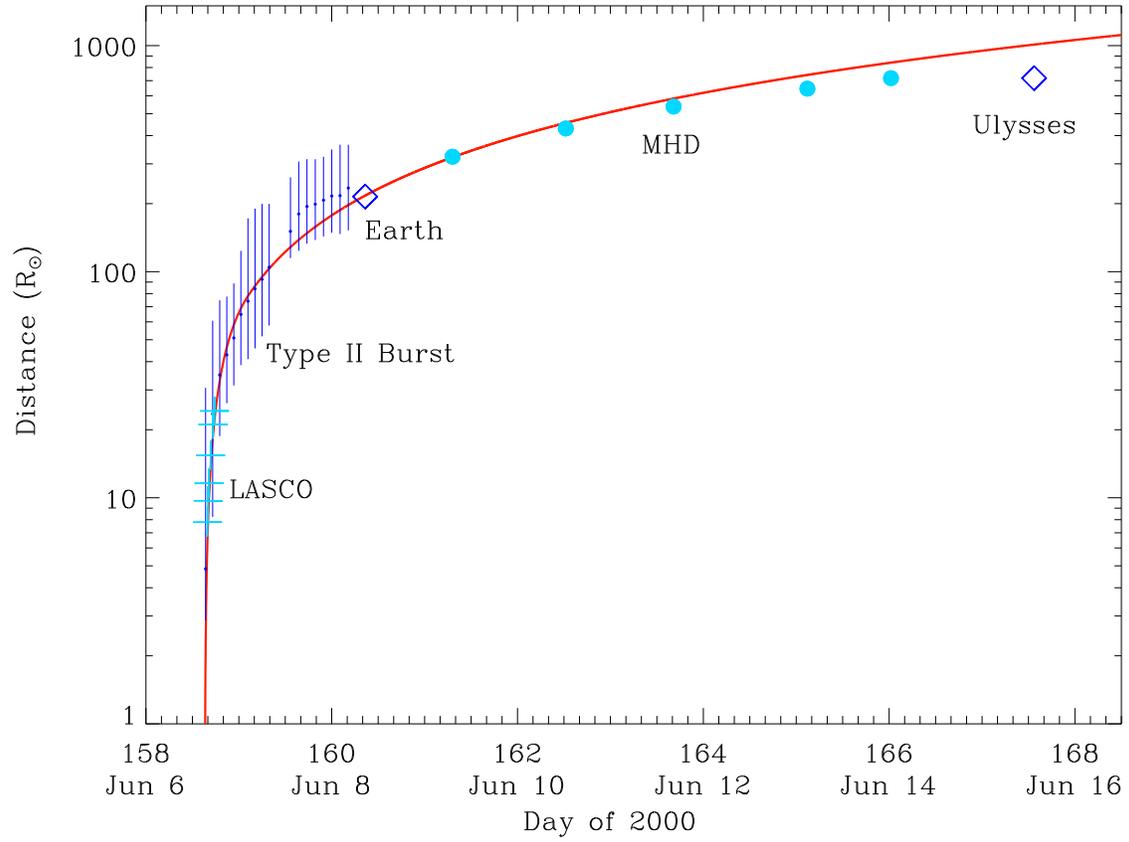}
\caption{\label{f10}Similar to Figure \ref{f5}, but for the overall propagation profile of the 2000 June 6 CME-driven shock from the Sun far into interplanetary space. Large blue dots show shock arrival times at 1.5, 2.0, 2.5, 3.0, and 3.34 au (\ulysses) predicted by the MHD model. }
\end{figure}

Figure \ref{f10} presents an extended propagation profile of the CME-driven shock from the analytical model. Also shown are distances obtained from the GCS modeling of the LASCO images, the frequencies of the observed type II radio bands, the locations of the Earth and \ulysses{} when the shock arrived there, and the MHD model output. The height--time curve is roughly consistent with all the data, but at larger distances from the Sun, it is higher than the MHD model output and the \ulysses{} location. Again, this is likely due to the large longitudinal and latitudinal separations between the Earth and \ulysses, which observed different solar wind backgrounds and shock structure, and the constant-speed assumption beyond 1 au in the analytical model. This height--time profile agrees with the assumption that the shock completes its major deceleration before reaching 1 au. The deceleration cessation distance (0.37 au) is smaller than the value given by \citet[0.46 au;][]{2007Reiner} and again falls in the range provided by \citet[0.2--0.4 au;][]{2013Liu} for three fast CMEs.

\subsection{The 2001 April 2 CME}

\begin{figure}
\epsscale{1.1} \plottwo{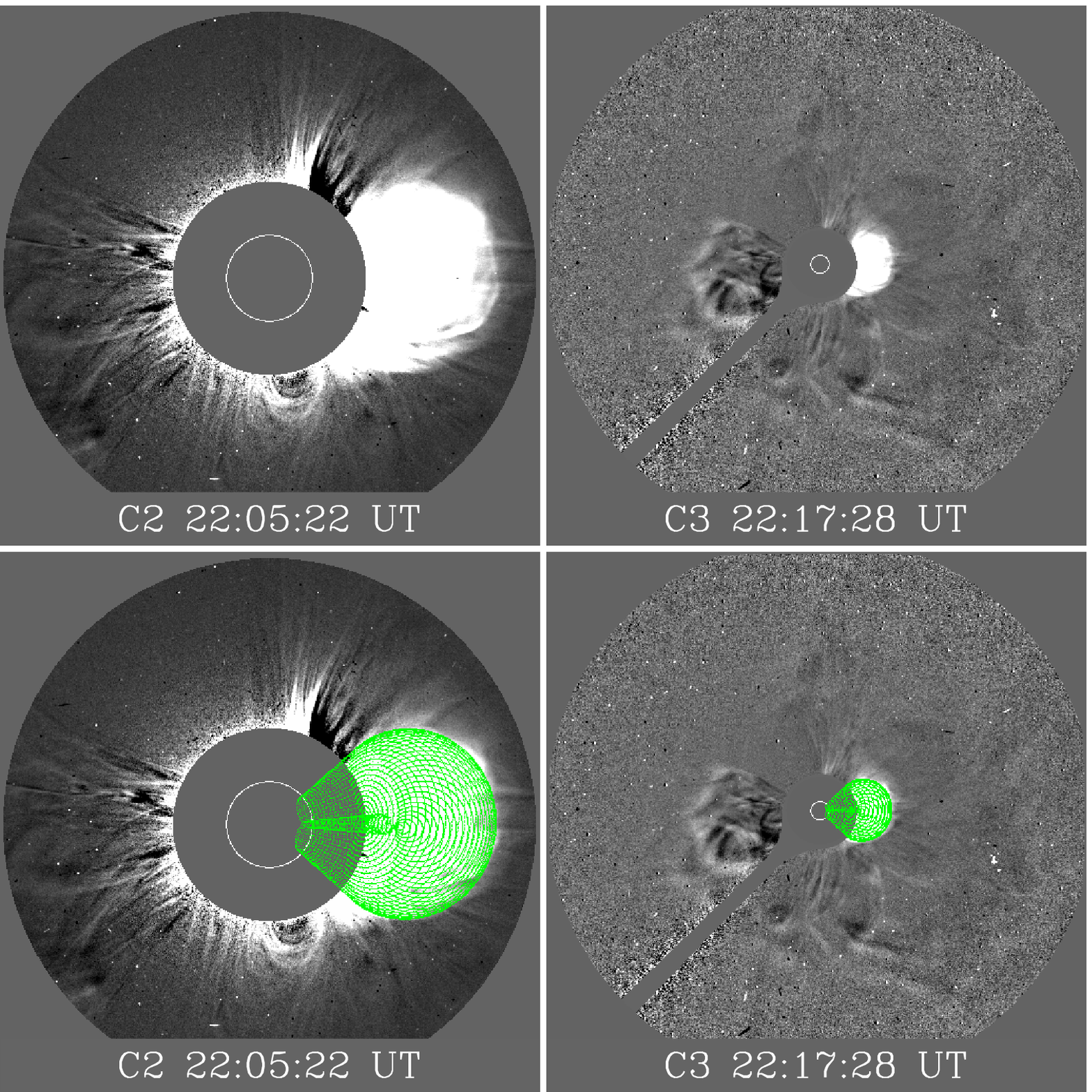}{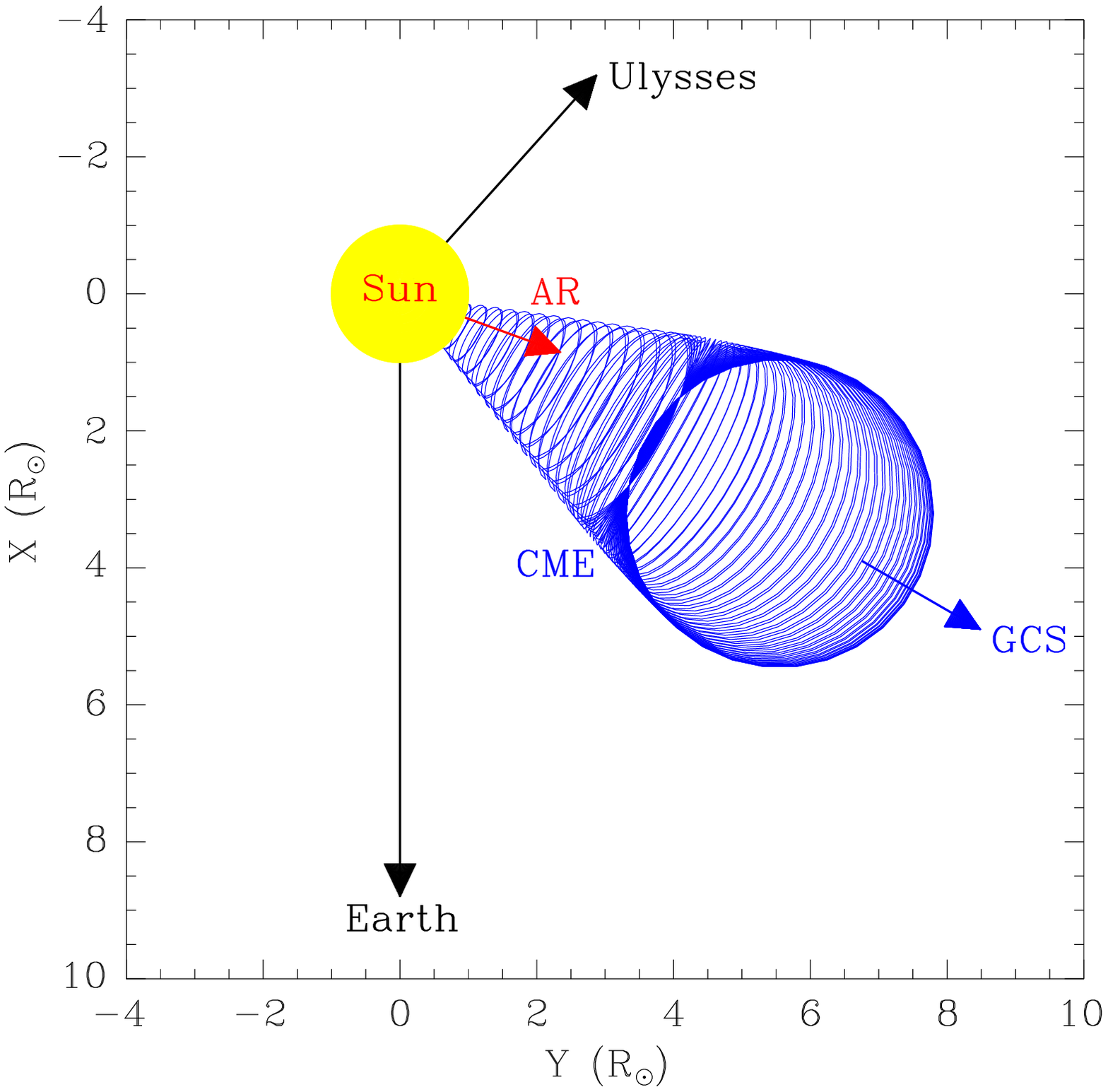}
\caption{\label{f11}Similar to Figure \ref{f1}, but for the 2001 April 2 CME. The modeled CME at 22:17:28 UT is projected onto the ecliptic plane. Ulysses was 26.2{\degr} south and 138.0{\degr} west of the Earth when the CME erupted from the Sun.  }
\end{figure}

On 2001 April 2, a CME originated from NOAA AR 09393 (N16{\degr}W70{\degr}) and was associated with an X20 flare peaking at 21:51 UT. Figure \ref{f11} (left) shows the coronagraph observations and the GCS wire frame rendering of the 2001 April 2 CME. This CME is a partial halo one preceded by several earlier CMEs. The CME at different times is fitted well by the GCS model. The longitude of the CME propagation direction near the Sun (W60{\degr}) agrees with its solar source longitude (W70{\degr}). The resulting latitude (S03{\degr}) indicates a small southward transition compared with the solar source latitude (N16{\degr}). The CME/shock speed near the Sun, estimated with a linear fit of the GCS heliocentric distances, is about 2300 \kms. The CME/shock may strike the Earth and \ulysses{} with its flanks given its propagation direction relative to the Earth and \ulysses{} (see Figure \ref{f11} right).

\begin{figure}
\epsscale{0.9} \plotone{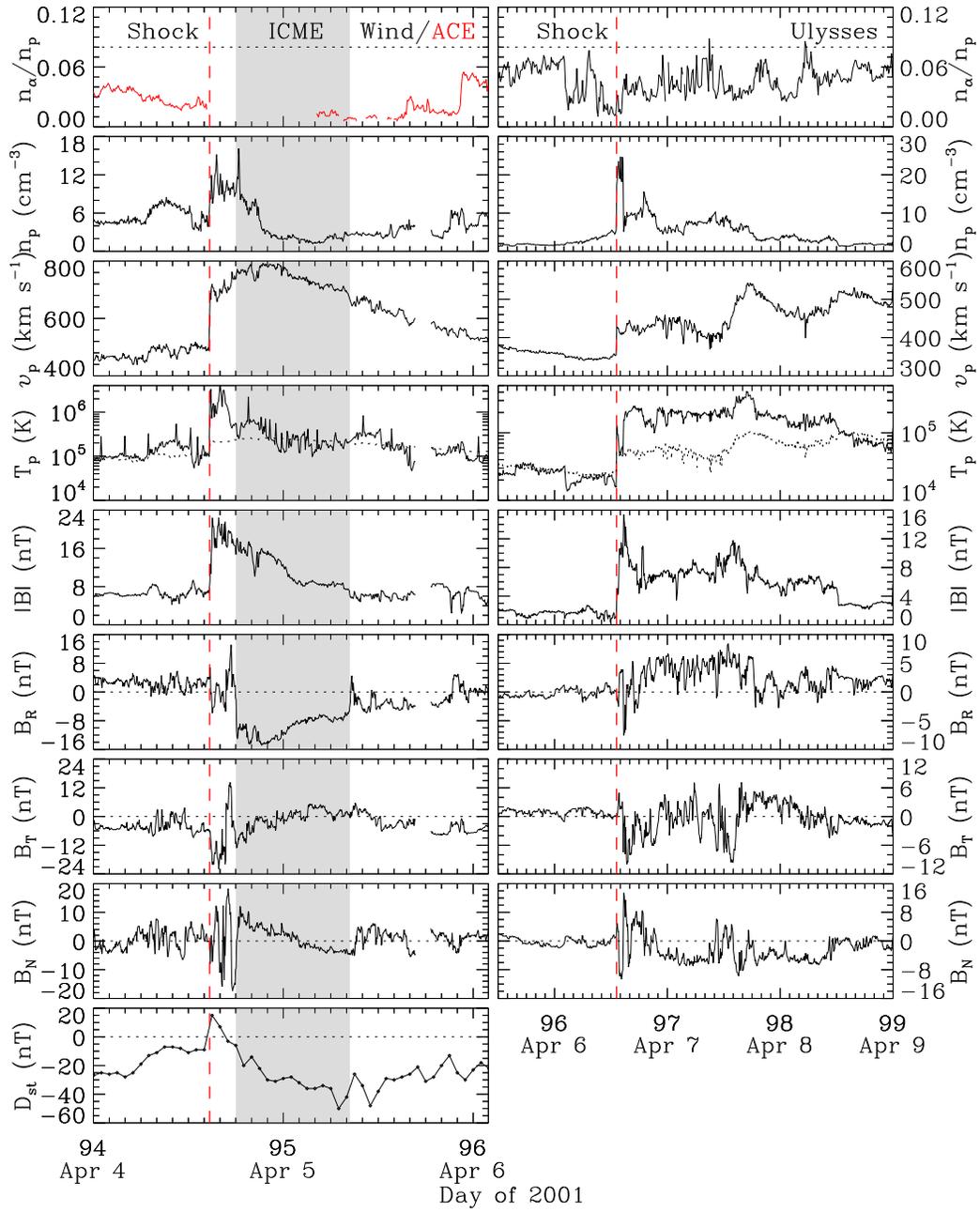}
\caption{\label{f12}Similar to Figure \ref{f7}, but for the in-situ measurements of the 2001 April 2 CME. The solar wind data at \wind{} (black) and \ace{} (red) are plotted in the left side of this figure.  }
\end{figure}

Figure \ref{f12} shows \wind{}/\ace{} and \ulysses{} in-situ solar wind measurements of the 2001 April 2 CME. A preceding shock was observed at \wind{} at about 14:41 UT on April 4 with a speed of about 870 \kms. The shaded region, a possible ICME interval, is identified mainly based on the slight rotation of the magnetic field components. The alpha-to-proton density ratio from \ace{} is minimal during the ICME. The Sun-to-Earth transit time of the shock ($\sim$41.0 hr) gives an average speed of about 1020 \kms. Although the southward magnetic field component in the sheath region reaches about ${-}$18 nT, it is very brief and no obvious geomagnetic storm was produced by the sheath field. The southward field component in the ICME trailing part triggered a small geomagnetic storm with a minimum \textit{D}$_\mathrm{st}$ index of ${-}$50 nT. A shock passed \ulysses{} at about 13:12 UT on April 6 with a speed of about 430 \kms. Ulysses was 23.8{\degr} south and 134.6{\degr} west of the Earth and at a distance of 1.45 au from the Sun. There are no clear ICME signatures following the shock at \ulysses. The in-situ measurements at the Earth and \ulysses{} are consistent with the encounter with the ICME flanks. The time duration from the Earth to \ulysses{} ($\sim$1.9 days) implies an average transit speed of about 410 \kms, which is similar to the shock speed at \ulysses. The shock observed at \ulysses{} might be the same as observed at the Earth. Comparison between the shock speed near the Sun (2300 \kms), the shock speed near the Earth (870 \kms), and the shock speed at \ulysses{} (430 \kms) suggests that the shock completed its major deceleration inside 1 au and afterwards it moved with a gradual deceleration.

\begin{figure}
\epsscale{0.8} \plotone{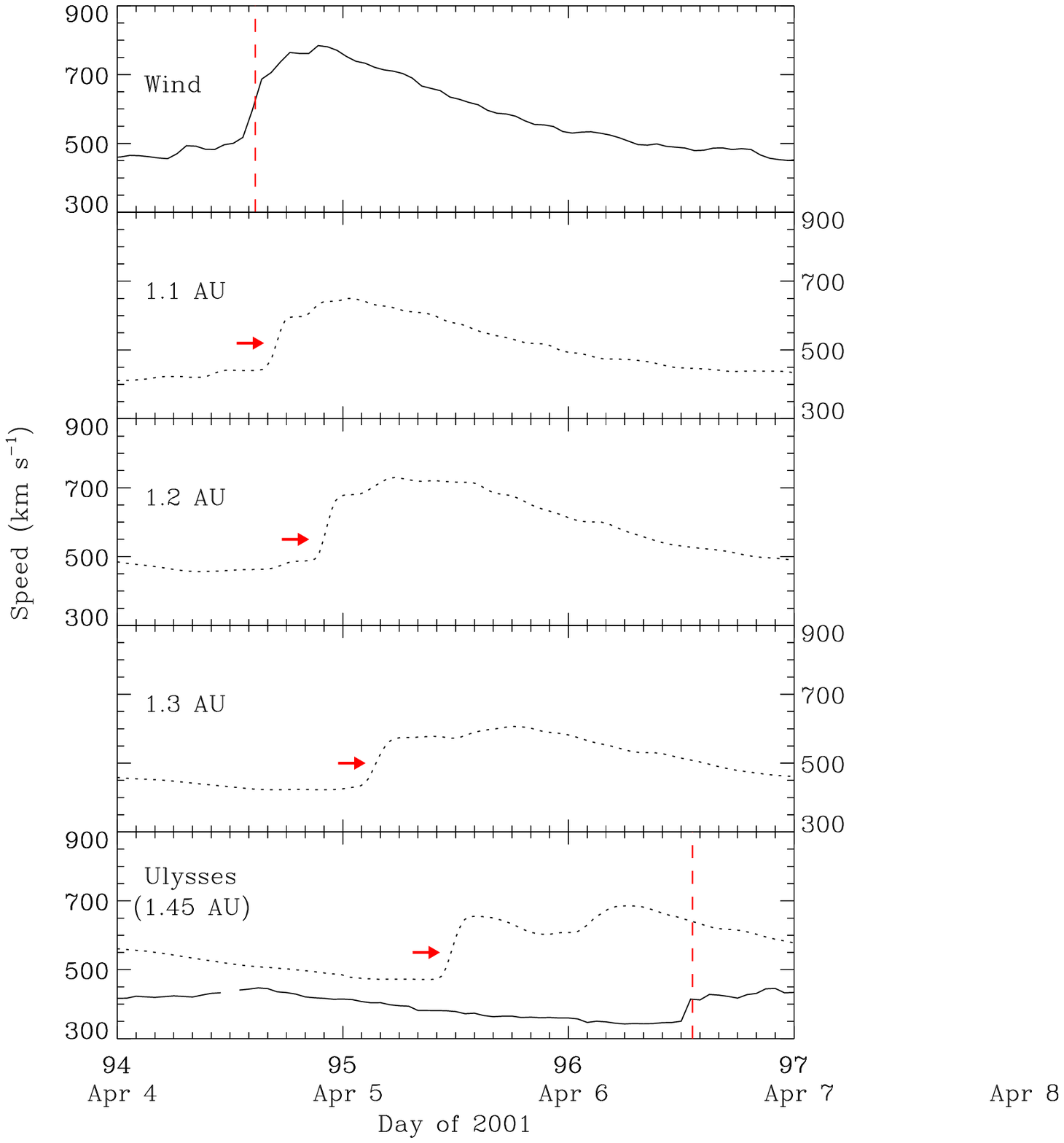}
\caption{\label{f13}Similar to Figure \ref{f3}, but for the solar wind speed evolution of the 2001 April 2 CME from \wind{} to \ulysses{} via the 1D MHD model.}
\end{figure}

The solar wind speeds at certain distances produced by the MHD model are presented in Figure \ref{f13}. The shock persists and is predicted to arrive at \ulysses{} at about 11:31 UT on April 5, which is about 26 hr earlier than observed. The time difference (26 hr) is still smaller than the transit time from the Earth to \ulysses{} ($\sim$46 hr), but indicates a larger uncertainty compared with other cases. This is probably due to the large Earth-Ulysses longitudinal separation and the variation in the shock structure. Note that the measured solar wind speeds at \ulysses{} are much smaller than the predicted speeds, which may attribute to different solar wind backgrounds at \ulysses{} than at the Earth.

\begin{figure}
\epsscale{1.0} \plotone{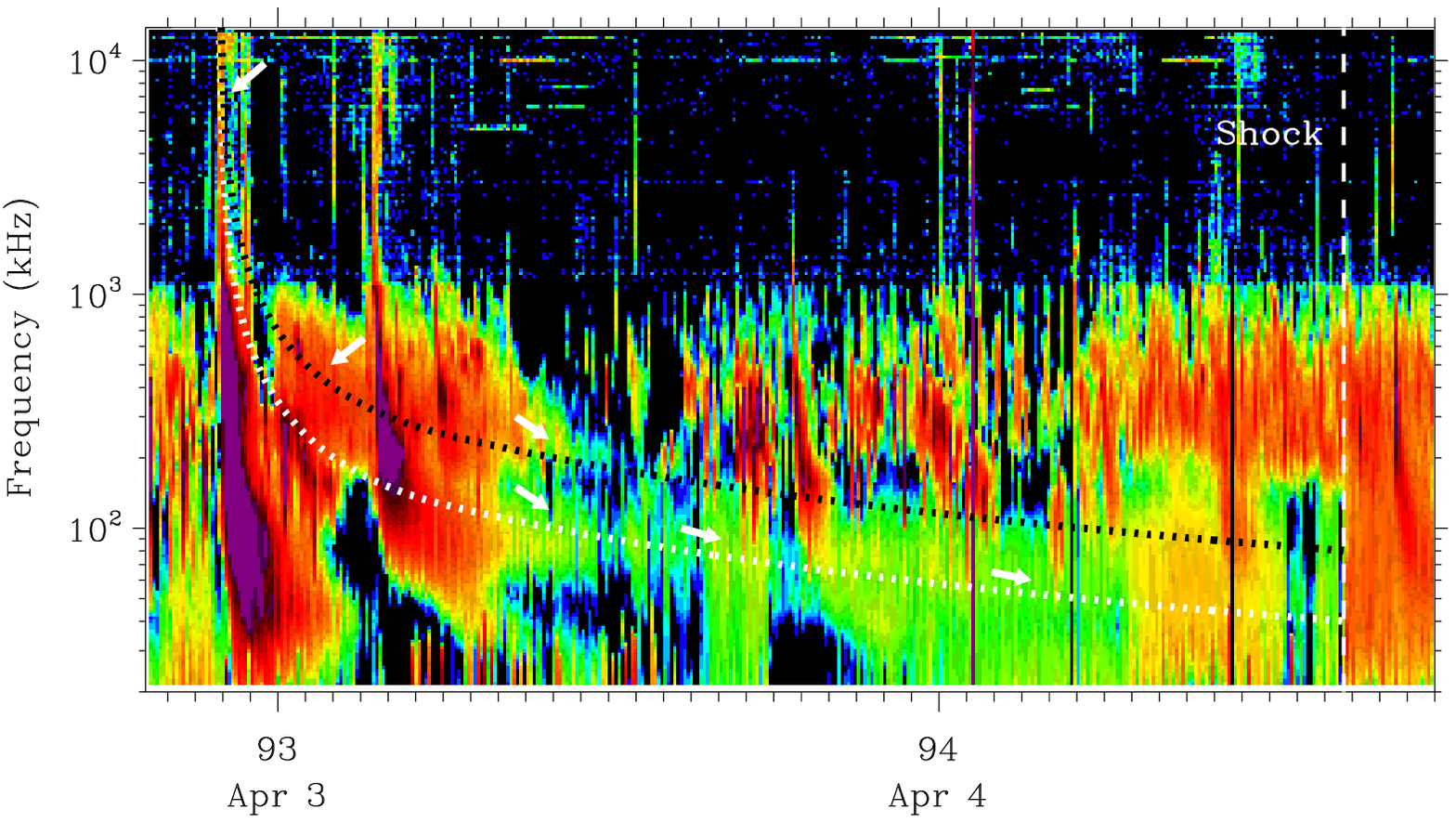}
\caption{\label{f14}Similar to Figure \ref{f4}, but for the dynamic spectrum associated with the 2001 April 2 CME. The white arrows indicate the observed type II radio bands.   }
\end{figure}

Figure \ref{f14} displays the radio dynamic spectrum associated with the 2001 April 2 CME. From about 22:00 UT to 22:24 UT on April 2, type II radio bands were observed at both the fundamental and harmonic plasma frequencies. An enhanced burst is observed, covering frequencies from about 180 kHz to 1000 kHz and times from about 23:45 UT on April 2 to 03:21 UT on April 3. We suggest that this special region is from the harmonic plasma frequency. After that the signatures of type II bursts are difficult to identify due to the contamination by type III emissions. After about 07:54 UT on April 3, the type II bursts may have reappeared and split into two weaker bands at the beginning and then drifted gradually at the fundamental plasma frequency until the shock arrived at the Earth. Again, caution should be taken for the identification of type II bursts as there are a series of type III bursts.

We input the shock initial speed near the Sun, the shock speed at the Earth and the Sun-Earth transit time into the analytical model, which gives a deceleration ${a}=-47.3$ m s$^{-2}$, a deceleration time period ${t}_{a}=8.4$ hr, and a deceleration cessation distance ${r}_{a}=0.32$ au. We convert the propagation distances from the analytical model to plasma frequencies based on the Leblanc density model \citep{1998Leblanc}. A scale value of ${n}_{0}=20$ cm$^{-3}$ yields a frequency-time profile that is consistent with the observed type II bands (see Figure \ref{f14}). In addition, the profile at the harmonic frequency also fits the enhanced region well.

\begin{figure}
\epsscale{1.0} \plotone{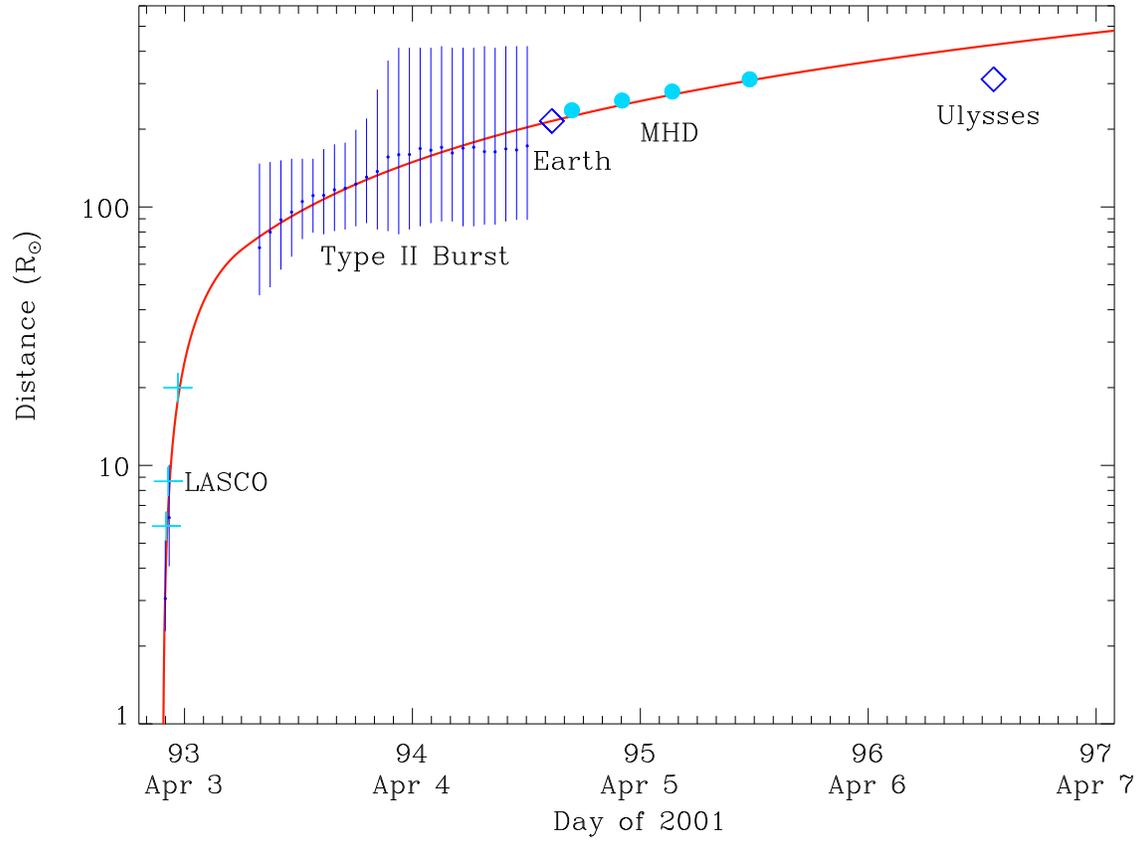}
\caption{\label{f15}Similar to Figure \ref{f5}, but for the overall propagation profile of the 2001 April 2 CME-driven shock. Large blue dots indicate shock arrival times at 1.1, 1.2, 1.3, and 1.45 au (\ulysses) predicted by the MHD model.  }
\end{figure}

The whole shock propagation profile from the Sun to the distance of \ulysses{} is shown in Figure \ref{f15}. The distances obtained from the GCS model, the type II radio bursts, in-situ measurements of the shock arrival at the Earth and \ulysses, and the MHD model results are also plotted in this figure. The height--time curve agrees well with all the data except the in-situ measurements at \ulysses, which may again result from the large longitudinal separation between the Earth and \ulysses{} and different solar wind backgrounds. The propagation profile again supports our assumption that the shock finishes its major deceleration before reaching 1 au and thereafter moves with a gradual deceleration. The shock deceleration cessation distance from the model (0.32 au) is similar to the result of \citet[0.38 au;][]{2007Reiner} for the same CME and also agrees with the results given by \citet[0.2--0.4 au;][]{2013Liu} for three fast CMEs.

\subsection{The 2001 November 4 CME}

\begin{figure}
\epsscale{1.1} \plottwo{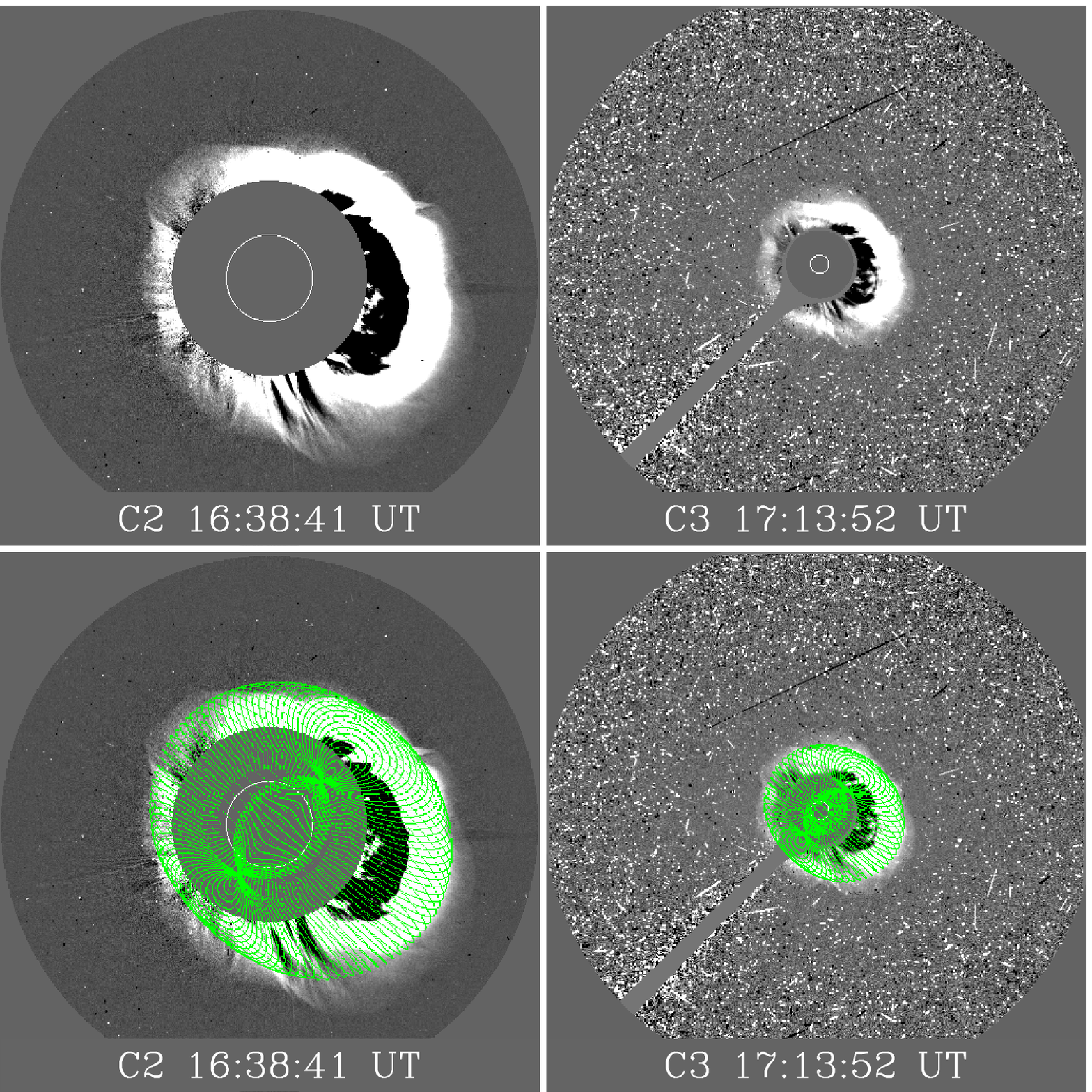}{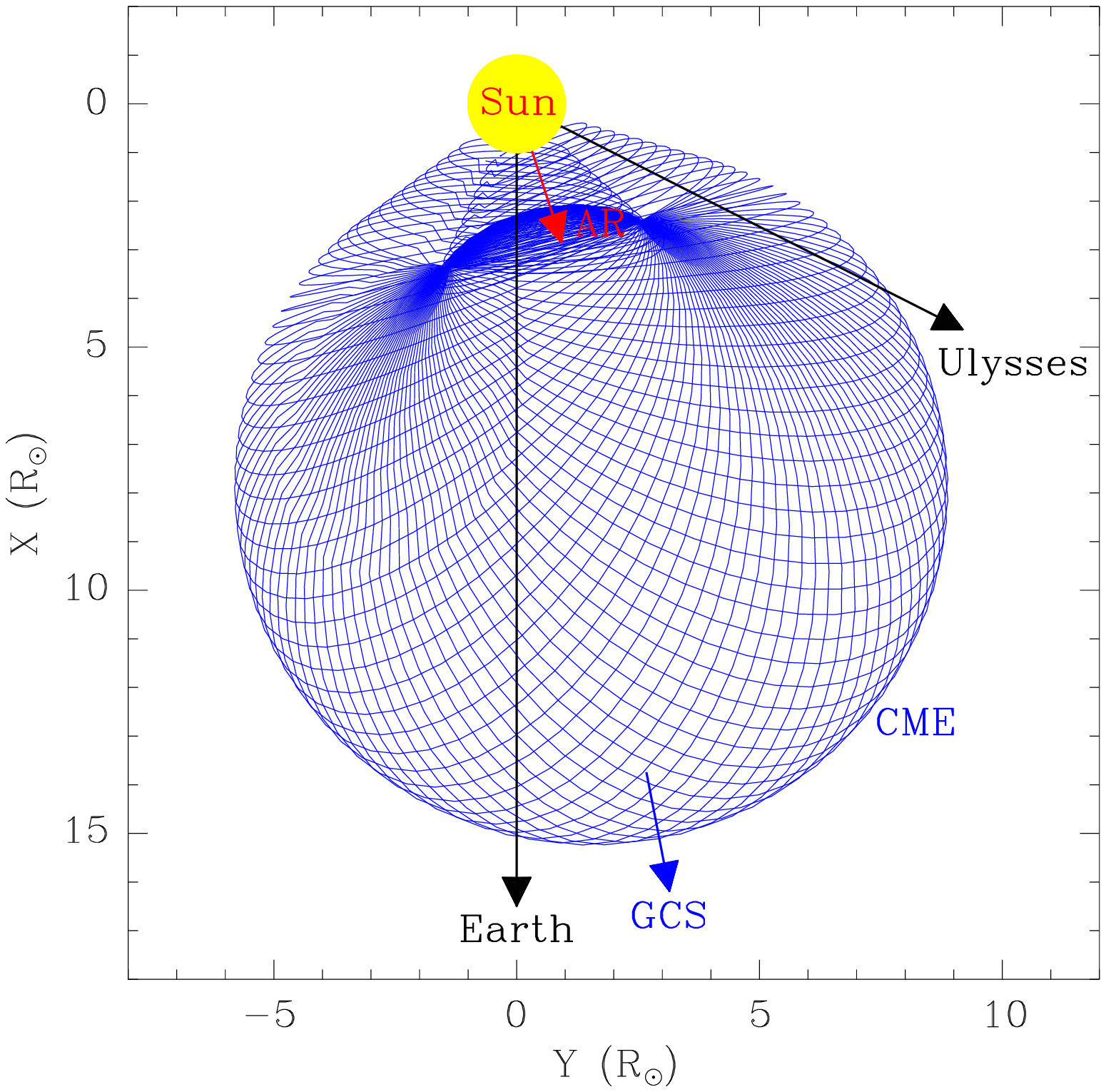}
\caption{\label{f16}Similar to Figure \ref{f1}, but for the 2001 November 4 CME. The modeled CME at 17:13:52 UT is projected onto the ecliptic plane. Note that \ulysses{} was 73.5{\degr} north and 63.2{\degr} west of the Earth when the CME launched from the Sun. }
\end{figure}

A halo CME erupted from NOAA AR 09684 (N06{\degr}W18{\degr}) on 2001 November 4 accompanied by an X1.0 flare peaking at 16:20 UT. Coronagraph observations and GCS modeling of the 2001 November 4 CME are displayed in Figure \ref{f16} (left). A shock is visible ahead of the CME front. The GCS model structure agrees well with the observed images. The CME propagates along 2{\degr} north and 11{\degr} west near the Sun, which is consistent with its solar source location (N06{\degr}W18{\degr}). The CME speed, derived from a linear fit of the heliocentric distances produced by the GCS model, is about 2600 \kms. This will be used as a proxy for the shock speed near the Sun. The CME was likely to impact \wind{} and \ulysses{} (see Figure \ref{f16} right). Note that \ulysses{} was close to the north pole.

\begin{figure}
\epsscale{0.9} \plotone{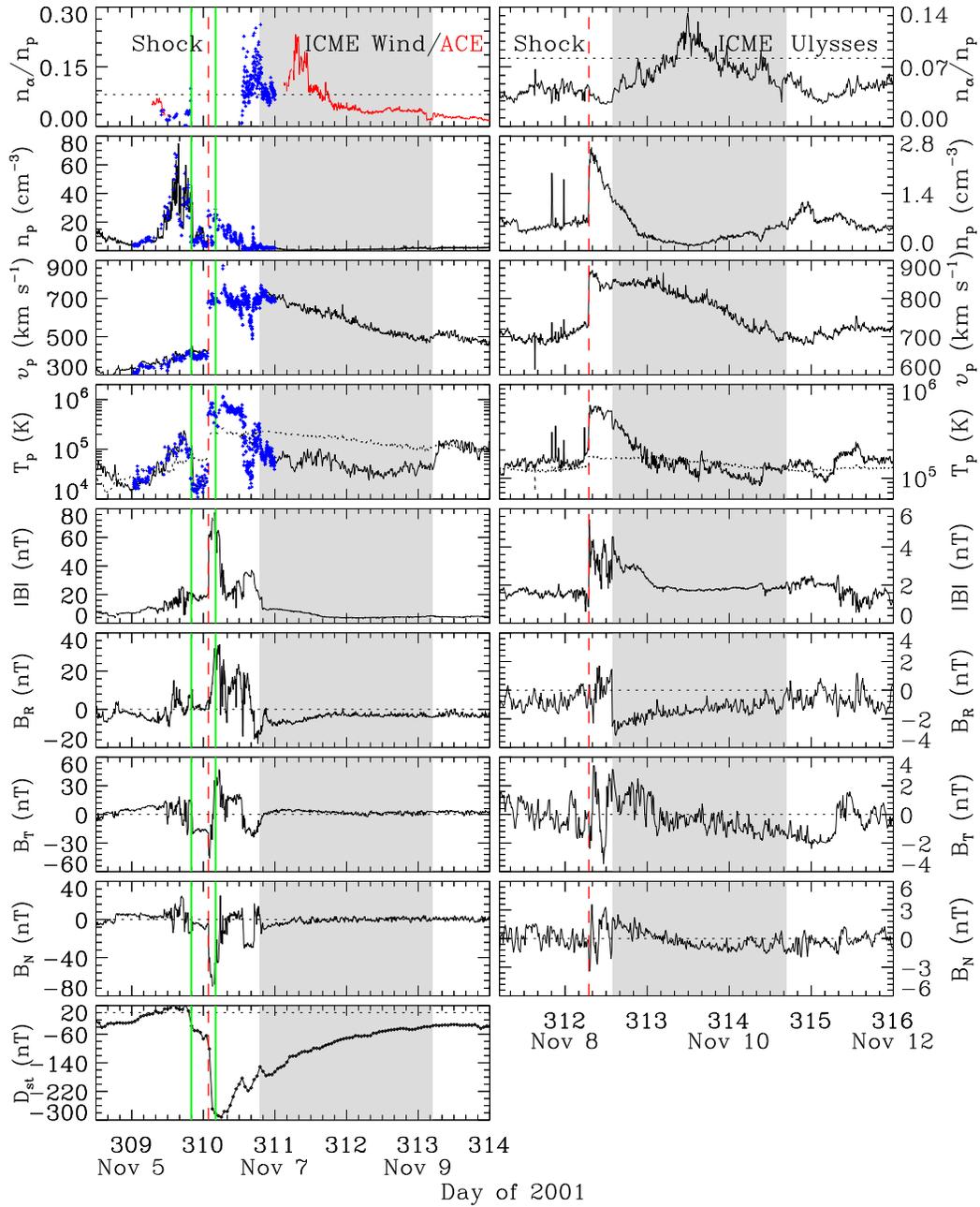}
\caption{\label{f17}Similar to Figure \ref{f12}, but for the in-situ measurements of the 2001 November 4 CME. Blue plus symbols (left) are solar wind data at \genesis. The two vertical green solid lines indicate the interval of a preceding ICME. }
\end{figure}

Figure \ref{f17} presents combined in-situ solar wind measurements of the 2001 November 4 CME at the near-Earth spacecraft and \ulysses. A strong forward shock passed \wind{} at about 01:40 UT on November 6 with a speed of about 1040 \kms. The shaded region indicates the identified ICME interval from 19:00 UT on November 6 to 04:48 UT on November 9. We determine the interval mainly using the depressed proton temperature and the declining speed profile. An earlier ICME (sandwiched by the two vertical green solid lines) is also recognized, which agrees with the results of \citet{2003Reisenfeld}. The preceding ejecta is accelerated, heated and compressed by the overtaking shock driven by the later CME, as in the events analyzed by \citet{2012Liu}. The CME/shock transit time from its eruption to the shock arrival at \wind{} is about 33.5 hr, which yields an average speed of about 1230 \kms. The southward magnetic field component is as large as about ${-}$70 nT in the sheath region, produced by the shock enhancement of the preexisting southward fields in the preceding ejecta. An intense complex geomagnetic storm with a minimum value of \textit{D}$_\mathrm{st}$ index of ${-}$292 nT occurred because of the enhanced southward magnetic field component in the sheath region and the high speed there.

Ulysses observed a shock at about 06:57 UT on November 8 at a distance of 2.21 au from the Sun. It was 73.2{\degr} north and 63.3{\degr} west of the Earth. An ICME is identified mainly from the rotation of the magnetic field components in combination with the enhanced alpha-to-proton density ratio. The preceding ejecta at the near-Earth spacecraft is not observed at \ulysses. The magnetic field signatures of the ICME at the Earth and \ulysses{} are not the same. They may have observed different parts of the ICME because of the large latitudinal and longitudinal separations between the two locations. The shock has a speed of about 920 \kms. Its transit time from the Earth to \ulysses{} ($\sim$2.2 days) gives an average speed of about 960 \kms. Comparison between the shock speed near the Sun (2600 \kms), the shock speed at the Earth (1040 \kms), and the shock speed at \ulysses{} (920 \kms) suggests that the major deceleration occurred inside 1 au and thereafter the shock underwent a gradual deceleration.

\begin{figure}
\epsscale{0.8} \plotone{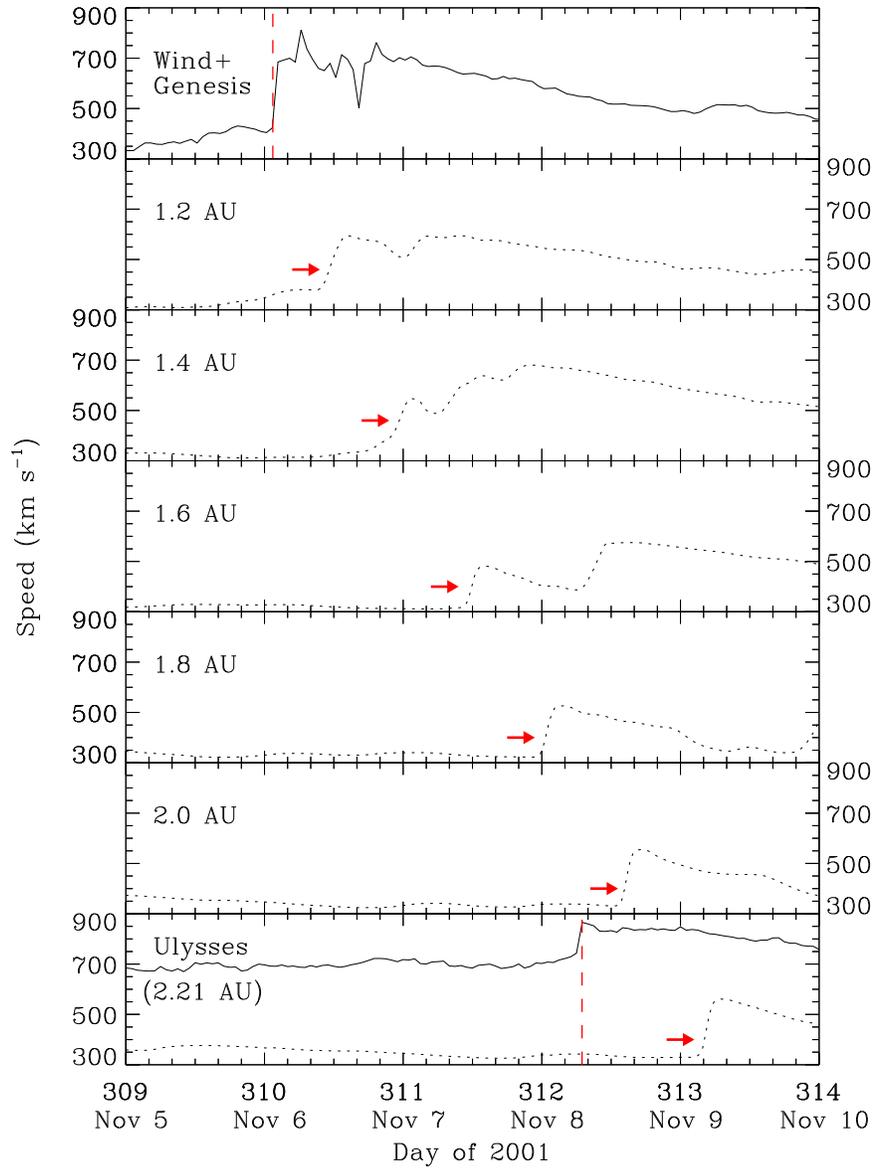}
\caption{\label{f18}Similar to Figure \ref{f3}, but for the solar wind speed evolution of the 2001 November 4 CME from the Earth to \ulysses{} via the 1D MHD model.}
\end{figure}

The MHD propagation results at certain distances are given in Figure \ref{f18}. The flow associated with the shock evolves into two streams, but the leading front should be used to track the shock. The shock persisted out to \ulysses, and the predicted shock arrival time at \ulysses{} is about 21.4 hr later than observed. The time lag (21.4 hr), which is likely caused by the large latitudinal and longitudinal separations between the Earth and \ulysses, is smaller than the shock transit time from the Earth to \ulysses{} ($\sim$53 hr) but cannot be neglected. The predicted solar wind at \ulysses{} is much slower than observed, which again indicates the different solar wind backgrounds. The higher solar wind speed at \ulysses{} is reasonable since \ulysses{} was near the north pole ($\sim$77{\degr}).

\begin{figure}
\epsscale{1.0} \plotone{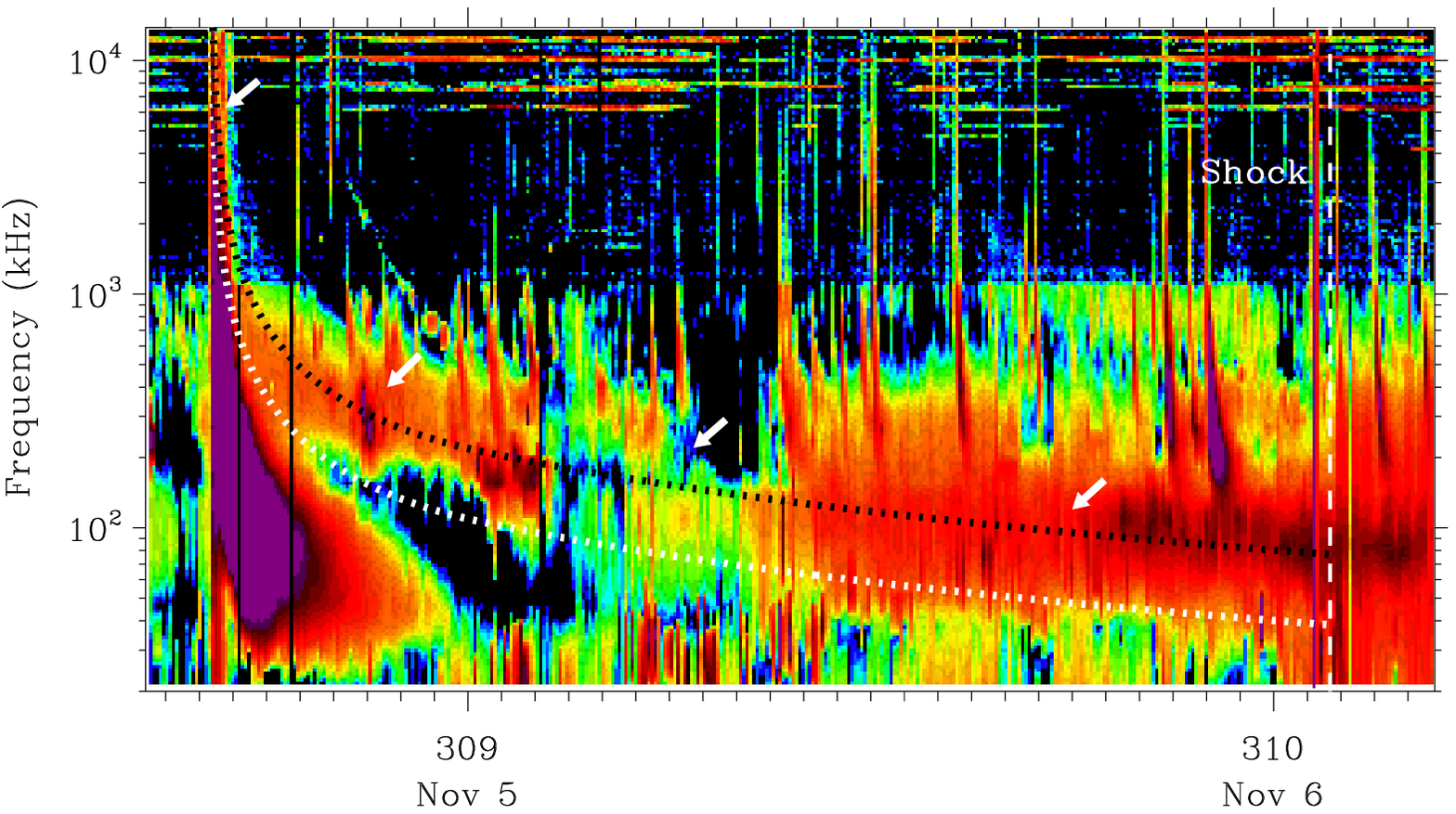}
\caption{\label{f19}Similar to Figure \ref{f4}, but for the dynamic spectrum associated with the 2001 November 4 CME. The white arrows indicate the observed type II radio bands. }
\end{figure}

An overview of the radio dynamic spectrum associated with the 2001 November 4 CME is presented in Figure \ref{f19}, which provides the link between the CME near the Sun and the ICME at 1 au. Diffuse but strong type II radio bands are mainly observed at the harmonic plasma frequency from about 16:28 UT on November 4 until the arrival of the shock at the Earth. With the insertion of the known parameters, the analytical model yields a deceleration ${a}=-52.6$ m s$^{-2}$, a time period for deceleration ${t}_{a}=8.2$ hr, and a deceleration cessation distance ${r}_{a}=0.36$ au. The resulting distances are then converted to plasma frequencies based on the Leblanc density model. A density scale of ${n}_{0}=18$ cm$^{-3}$ gives a frequency-time profile that is well consistent with the type II bands.

\begin{figure}
\epsscale{1.0} \plotone{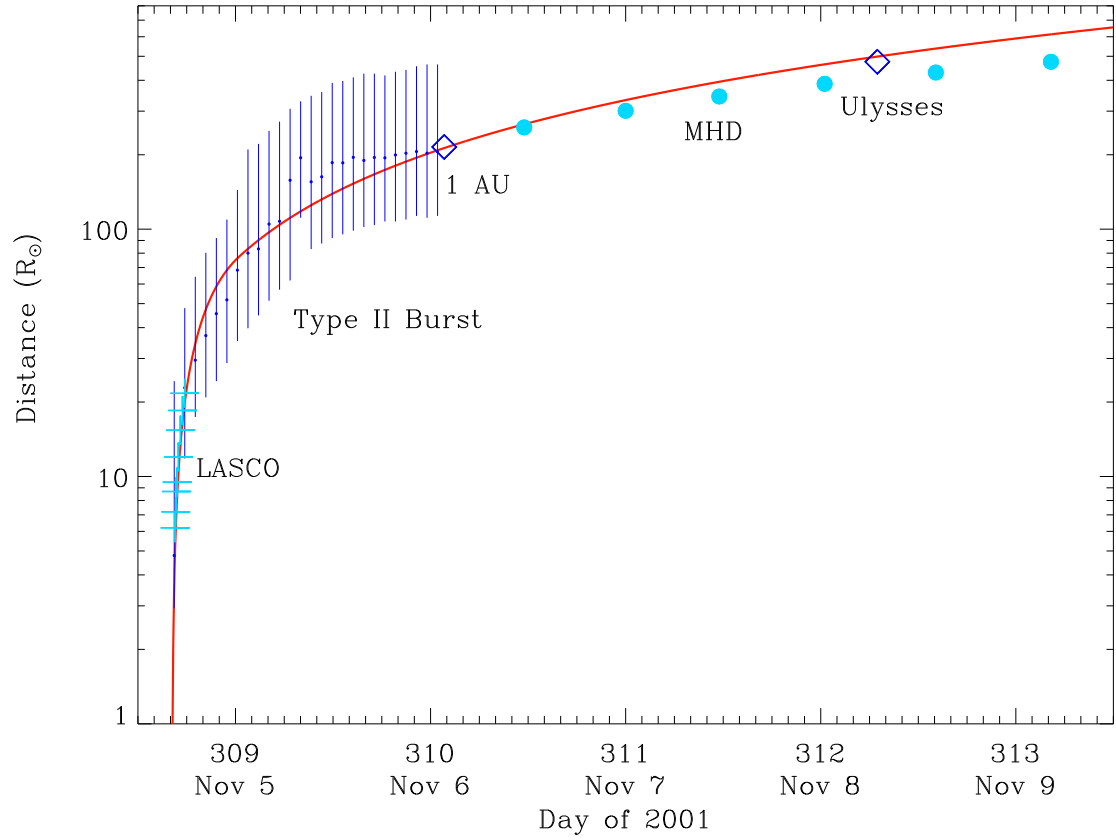}
\caption{\label{f20}Similar to Figure \ref{f5}, but for the overall propagation profile of the 2001 November 4 CME-driven shock. Large blue dots represent the shock arrival times at 1.2, 1.4, 1.6, 1.8, 2.0, and 2.21 au (\ulysses) predicted by the MHD model.}
\end{figure}

Figure \ref{f20} shows the shock propagation profile from the Sun to beyond the Earth. The analytical profile is consistent with the distances from the GCS model, the averaged distances derived from the observed type II bands, and the shock arrival at the Earth and \ulysses. The MHD model results are consistent with the analytical profile first but become gradually lower than the profile. Again, this is owing to the different solar wind backgrounds at the Earth and \ulysses{} in combination with the constant-speed assumption beyond 1 au in the analytical model. In general, the simple analytical model of shock propagation, in particular the assumption that the shock finishes its major deceleration before it arrives at 1 au, is verified by the agreement. The deceleration cessation distance of the shock (0.36 au) agrees with results of \citet[0.2--0.4 au;][]{2013Liu}. The shock speed might have been changed due to its interaction with the preceding ejecta, although it is not clear when the interaction began (see Figure \ref{f17}).

\subsection{The 2001 November 22 CME}

\begin{figure}
\epsscale{0.6} \plotone{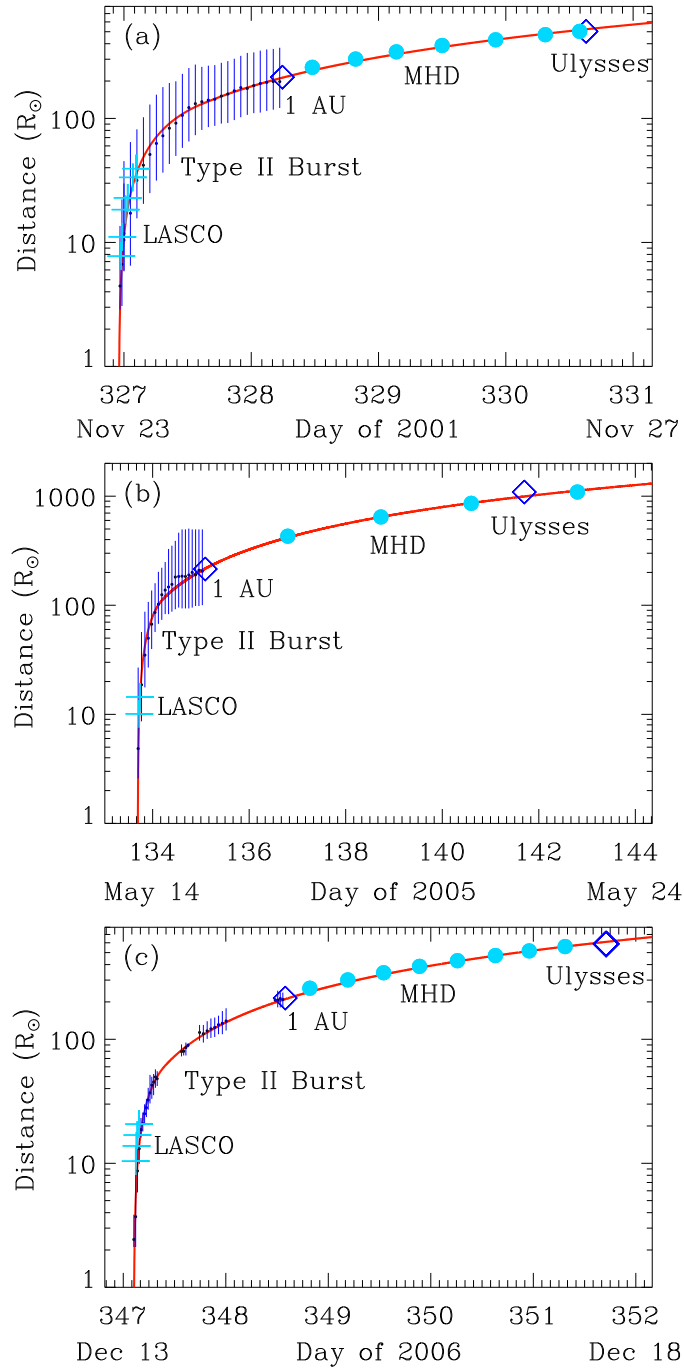}
\caption{\label{f21}Similar to Figure \ref{f5}, but for the overall propagation profiles of the 2001 November 22 CME (a), the 2005 May 13 CME (b), and the 2006 December 13 CME (c). Reproduced from \citet{2017Liu}, \citet{2017Zhao} and \citet{2008LiuL}, respectively. }
\end{figure}

On 2001 November 22, a CME erupted from NOAA AR 09704 (S17{\degr}W36{\degr}) associated with an M9.9 flare peaking at 23:30 UT. \citet{2017Liu} provide a comprehensive view of the evolution of the 2001 November 22 CME/shock from the Sun to deep interplanetary space. In their studies, a complex type II radio burst plays a key role in determining the shock propagation and its interaction with earlier CMEs. They find that the shock decelerates rapidly in the Sun-Earth space and then moves with a roughly constant speed or gradual deceleration by employing the analytical model (see Section \ref{prop}) based on coordinated coronagraph and multi-point in-situ observations. All the data associated with the shock propagation in their study are shown in Figure \ref{f21}(a). The shock is tracked very well from the Sun to the distance of \ulysses{}, and such good agreement verifies the analytical model with the assumption that the shock finishes its major deceleration inside 1 au. We use the propagation parameters obtained by \citet{2017Liu} for our statistical analysis, i.e., a deceleration ${a}=-20.8$ m s$^{-2}$, a time for deceleration ${t}_{a}=15.4$ hr, and a deceleration cessation distance ${r}_{a}=0.60$ au. \citet{2017Liu} attribute the large deceleration cessation distance to the shock propagation inside a preceding CME (see their paper for details).

\subsection{The 2005 May 13 CME}

On 2005 May 13, a CME erupted from NOAA AR 10759 (N12{\degr}E11{\degr}). It was accompanied by a long-duration M8.0 flare, which peaked at 16:57 UT. The propagation of the 2005 May 13 CME/shock from the Sun to the distance of \ulysses{} has been studied by \citet{2017Zhao}. They employ a kinematic model as in \citet{2008LiuL} to fit the frequency drift of the observed type II radio burst. Now we provide a new analysis of the CME/shock propagation by applying the analytical model without free parameters (see Section \ref{prop}). With the input of the shock speed near the Sun (2500 \kms), the shock transit time (33.6 hr), and the shock speed at the Earth (950 \kms), the analytical model gives a deceleration ${a}=-32.8$ m s$^{-2}$, a time for deceleration ${t}_{a}=13.1$ hr, and a deceleration cessation distance ${r}_{a}=0.54$ au. The new shock propagation profile and all the data set from \citet{2017Zhao} are given in Figure \ref{f21}(b). The new profile accords well with all the data although it is slightly lower than the distance of \ulysses. Note that the CME-driven shock missed \ulysses, and the ICME leading edge has been used as a substitute. The shock propagation parameters obtained from the analytical model are similar to those from \citet{2017Zhao}. Again, the shock completes its major deceleration inside 1 au.

\subsection{The 2006 December 13 CME}

On 2006 December 13, a CME was launched from NOAA AR 10930 (S06{\degr}W23{\degr}) and associated with an X3.4 flare peaking at 02:40 UT. A kinematic model with two free parameters is used by \citet{2008LiuL} to fit the frequency drift of the associated type II burst. In this work, we perform a new analysis using the analytical model with no free parameters (see Section \ref{prop}). Based on LASCO coronagraph observations, we reconstruct the CME with the GCS method (not shown here). The estimated CME propagation direction near the Sun (S08{\degr}W03{\degr}) accords with its solar source location (S06{\degr}W23{\degr}). The CME/shock speed near the Sun from a linear fit of the GCS distances is about 2200 \kms, which is almost the same as the shock speed obtained by \citet[2212 \kms;][]{2008LiuL}. The shock parameters are inserted into the analyitcal model, which gives a deceleration ${a}=-44.2$ m s$^{-2}$, a time for deceleration ${t}_{a}=7.3$ hr, and a deceleration cessation distance ${r}_{a}=0.28$ au. Figure \ref{f21}(c) presents the new propagation profile of the shock from the Sun to the distance of \ulysses. Distances produced by the GCS model and other relevant data given by \citet{2008LiuL} are also plotted in this figure. The new profile agrees well with all the data, which comfirms the simple analytical model once again. The new deceleration cessation distance (0.28 au) is slightly smaller than that of \citet[0.36 au;][]{2008LiuL}. Again, the shock finished its major deceleration within 1 au.

\section{Statistical Analysis}\label{stat}

The kinematic parameters of the 7 events are given in Table \ref{propa}, including the Sun-Earth shock transit time, the shock speed near the Sun, the shock speeds at the Earth and \ulysses{}, and the deceleration parameters from the analytical model. Here, we investigate their correlations in order to have a statistical sense of CME/shock propagation characteristics in interplanetary space. A first impression from Table \ref{propa} by comparing the shock speeds near the Sun, at the Earth and at \ulysses{} is that the events finished their major deceleration within 1 au and thereafter moved with a roughly constant speed or gradual deceleration. This verifies the assumption of the analytical model.

\begin{figure}
\epsscale{1.2} \plotone{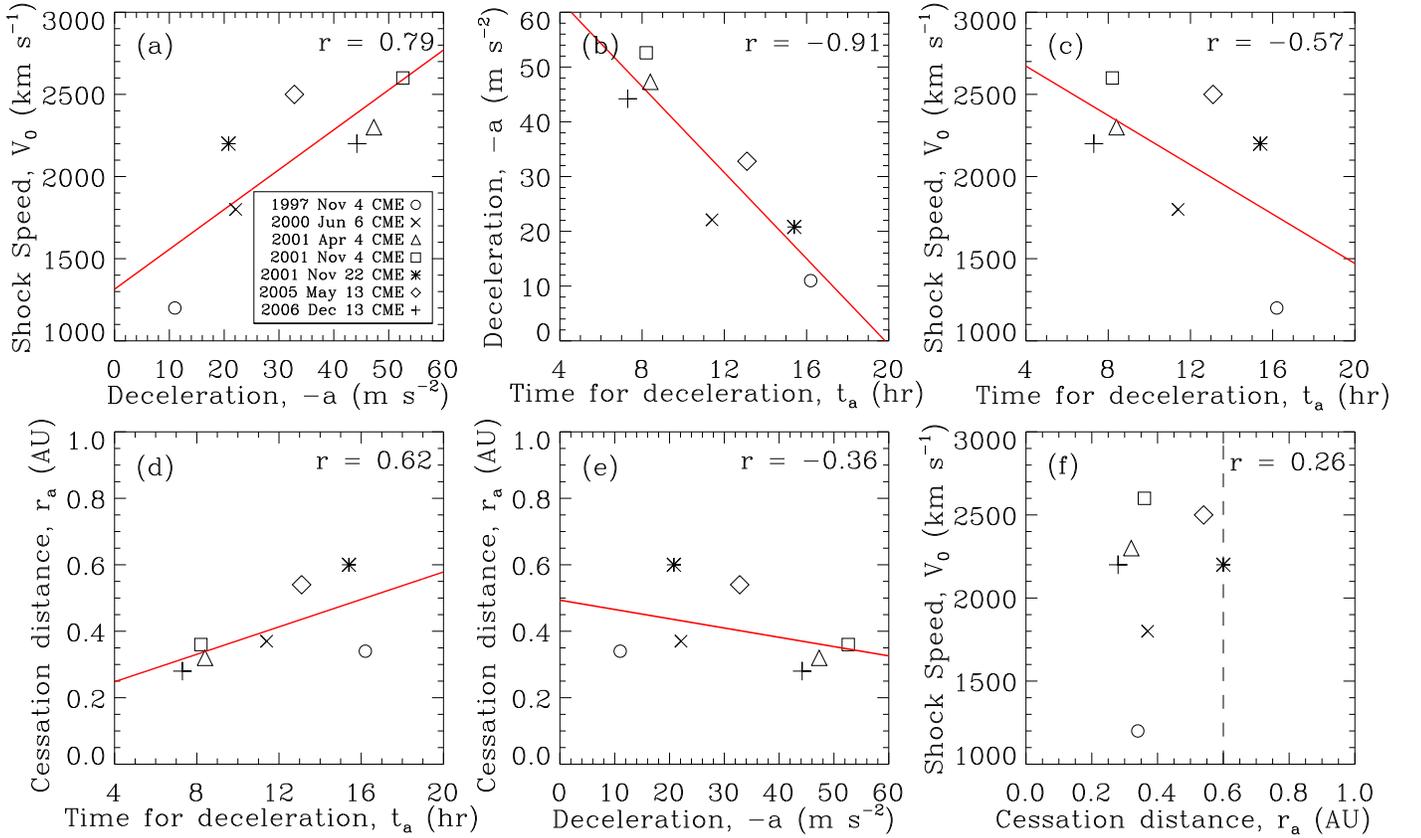}
\caption{\label{f22} Correlations of kinematic parameters generated by the analytical model for the 7 events. (a) Shock speed near the Sun versus the shock deceleration. (b) Shock deceleration versus the deceleration time period. (c) Shock speed near the Sun versus the deceleration time period. (d) Shock deceleration cessation distance versus the shock deceleration time period. (e) Shock deceleration cessation distance versus the shock deceleration. (f) Shock speed near the Sun versus the deceleration cessation distance. The red lines indicate the linear fit results for each pair. The vertical dashed line in the last panel marks the distance of 0.60 au from the Sun. The corresponding correlation coefficient is given in each panel. Different symbols represent different CMEs (see the first panel).}
\end{figure}

Figure \ref{f22}(a) displays the shock speed near the Sun versus the shock deceleration for the 7 events. A CME/shock with a higher initial speed tends to have a larger deceleration. The linear correlation coefficient between the shock initial speed and its deceleration is about 0.79. This suggests that faster events decelerate more rapidly within 1 au. Figure \ref{f22}(b) shows the shock deceleration versus the time period of shock deceleration for the 7 events. The correlation coefficient is about ${-}$0.91, which indicates that a CME/shock with a larger deceleration tends to have a shorter deceleration time period from the Sun to 1 au. It is surprising that the shock initial speed is roughly anti-correlated with the time period of shock deceleration for the 7 events, which can be seen in Figure \ref{f22}(c). The linear correlation coefficient between them is about ${-}$0.57, which implies that a faster event tends to decelerate for a shorter time period during its transit in the Sun-Earth space. Figure \ref{f22}(d) gives the relationship between the shock deceleration cessation distance and the deceleration time period for the 7 events with a linear correlation coefficient of $\sim$0.62: a CME/shock that decelerates for a shorter time period also tends to have a shorter cessation distance. A weak correlation between the shock deceleration cessation distance and the shock deceleration is shown in Figure \ref{f22}(e) with a correlation coefficient only about ${-}$0.36. Figure \ref{f22}(f) shows no obvious correlation between the shock initial speed and the deceleration cessation distance for the 7 events. The sample may be small, so a definite correlation between them cannot be obtained. All the deceleration cessation distances are no larger than 0.60 au, which, again, agrees with the assumption that a CME/shock finishes its major deceleration before reaching 1 au. These correlations in Figure \ref{f22} are consistent with the results of \citet{2007Reiner} and \citet{2016Liu} that faster CMEs/shocks tend to decelerate more rapidly and for shorter time periods. It should be stressed that the CMEs we have studied are all fast events with speeds near the Sun above 1000 \kms.

\section{Conclusions and Discussion}\label{sum}

We have investigated the propagation and evolution of 7 fast CMEs/shocks from the Sun into interplanetary space beyond 1 au, combining coronagraph observations, type II radio bursts and in-situ measurements at the Earth and \ulysses. This work demonstrates how to quantify the propagation of CMEs/shocks in the heliosphere based on a simple but effective analytical model, which is important for space weather research and forecasting. We then perform a correlation analysis of the kinematical parameters of the 7 events in order to have a statistical sense of CME/shock propagation behavior in interplanetary space. The results are summarized and discussed below.

By comparing the shock speed near the Sun, the shock speed at the Earth, and finally the shock speed at \ulysses, we find that each of the 7 CME-driven shocks finished its major deceleration before reaching 1 au and thereafter moved with a gradual deceleration. This agrees with the Sun-Earth propagation profile of a typical fast CME discovered by \citet{2013Liu}, i.e., a rapid deceleration within 1 au followed by a nearly constant speed or gradual deceleration. The application of the analytical model to all events also gives a range of deceleration cessation distances from 0.28 au to 0.60 au, which further proves that the shock ceases its rapid deceleration in the Sun-Earth space. Correlations between the shock speed near the Sun, the shock deceleration, and the time period for shock deceleration suggest that a CME/shock with a higher speed near the Sun tends to have a larger deceleration and a shorter time period for the deceleration. This agrees with the results of \citet{1985Woo}, \citet{2007Reiner} and \citet{2016Liu}. Weak or no correlations are obtained between the deceleration cessation distance and the deceleration and between the shock speed near the Sun and the deceleration cessation distance.

Our analysis of the 7 fast events verifies the effectiveness and ease of use of the simple analytical model proposed by \citet{2017Liu} for the propagation of CMEs/shocks from the Sun to a large distance beyond 1 au. The analytical model has no free parameters and has been applied to each of the 7 events. The resulting propagation profile for each of the 7 events matches the distances obtained from the GCS fit near the Sun and the frequency drift of the type II radio burst between the Sun and the Earth, and is roughly consistent with the MHD model results beyond 1 au and the shock arrival time at \ulysses. This agreement further supports that the major deceleration of a CME/shock is finished inside 1 au. It also shows that the analytical model is a useful tool for CME research and space weather forecasting.

In this work, the deceleration of CMEs/shocks in the Sun-Earth space can be classified into two types. First, the deceleration for 5 of the 7 events is mainly caused by the interaction with the ambient solar wind, which confirms that CMEs tend to move together with the ambient solar wind before arriving at 1 au \citep[e.g.,][]{1999Lindsay, 2001GopalswamyL, 2010Maloney, 2016Liu}. The deceleration cessation distances for the 5 events (including possibly the 1997 November 4, the 2000 June 6, the 2001 April 2, the 2005 May 13, and the 2006 December 13 CMEs) range from 0.28 au to 0.54 au. The slightly larger cessation distance (0.54 au) for the 2005 May 13 CME may be owing to a different coronal and solar wind background. Second, the deceleration for 2 events is likely affected by the interaction with earlier CMEs in addition to the drag from the ambient solar wind. This may enlarge the deceleration cessation distance, for example 0.60 au for the 2001 November 22 CME. As \citet{2017Liu} suggest, the long-time interaction between the November 22 event and a preceding CME is likely responsible for the large deceleration cessation distance. Note that this is different from the preconditioning effect found by \citet{2014LiuL}. The earlier CMEs occurred either too early or too late. We compare the shock speed near the Sun and that at the Earth for each of the 7 events and find that they all had a significant deceleration within 1 au. Therefore, there may not be an obvious preconditioning effect for the events studied here.

Finally, we note that several intense geomagnetic storms occurred among the 7 events. The occurrence of the geomagnetic storms stresses the importance of the enhanced southward magnetic fields in the sheath region between the shock and ejecta. For the 2001 November 4 CME, a severe complex geomagnetic storm with a minimum \textit{D}$_\mathrm{st}$ index of ${-}$292 nT occurred resulting from the interaction between the shock and a preceding ejecta. These results are consistent with previous findings on the generation of large geomagnetic storms by ICME sheaths and shock-ejecta interaction \citep[e.g.,][]{1987Burlaga, 1992Tsurutani, 2017Kilpua, 2017Liu, 2017Lugaz}.

\acknowledgments

This research was supported by NSFC under grants 41774179 and 41604146 and the Specialized Research Fund for State Key Laboratories of China. We acknowledge the use of data from \textit{SOHO}, \wind{}, \ace{}, \textit{Genesis} and \ulysses{} and the \textit{D}$_\mathrm{st}$ index from WDC in Kyoto. The \textit{SOHO} data are available at \url{ftp://lasco6.nascom.nasa.gov/pub/lasco_level05/}, and the in-situ data are obtained from \url{ftp://spdf.gsfc.nasa.gov/pub/data/}.

\clearpage

\begin{table}
\tabcolsep=0.11cm
\caption{Solar source information, CME propagation direction, shock arrival time and \ulysses{} position relative to the Earth for the 7 CMEs. \label{tab}}
\begin{tabularx}{\linewidth}{@{\extracolsep{\fill}}ccccccccc}
\hline\hline
Date & AR & Flare & Direction\tablenotemark{a} & Shock Time\tablenotemark{b} & Shock Time\tablenotemark{c} & Lat\tablenotemark{d} & Lon\tablenotemark{d} & Dist\tablenotemark{d} \\
  &  &  &  & (UT)  & (UT)  & $(^{\circ})$  & $(^{\circ})$  & (au) \\
\hline
\tablenotemark{1}1997 Nov 4                    & 08100 (S14W33) & X2.1   & S04W29  & Nov 6, 22:19   & Nov 24, 15:07  & S2.0   & W114.6   & 5.33 \\
\tablenotemark{2}2000 Jun 6                    & 09026 (N20E18) & X2.3   & N11E03  & Jun 8, 08:38   & Jun 15, 13:26  & S57.4  & E83.0    & 3.39 \\
\tablenotemark{3}2001 Apr 2                    & 09393 (N16W70) & X20    & S03W60  & Apr 4, 14:41   & Apr 6, 13:12   & S26.2  & W138.0   & 1.46 \\
\tablenotemark{4}2001 Nov 4                    & 09684 (N06W18) & X1.0   & N02W11  & Nov 6, 01:40   & Nov 8, 06:57   & N73.5  & W63.2    & 2.18 \\
\tablenotemark{5}2001 Nov 22\tablenotemark{e}  & 09704 (S17W36) & M9.9   & N08W12  & Nov 24, 05:53  & Nov 26, 15:07  & N71.9  & W60.1    & 2.31 \\
\tablenotemark{6}2005 May 13\tablenotemark{f}  & 10759 (N12E11) & M8.0   & S01E02  & May 15, 02:09  & missing        & S19.8  & E71.4    & 5.11 \\
\tablenotemark{7}2006 Dec 13\tablenotemark{g}  & 10930 (S06W23) & X3.4   & S08W03  & Dec 14, 14:00  & Dec 17, 17:02  & S72.9  & W114.1   & 2.76 \\
\hline
\end{tabularx}
\tablenotetext{a}{Estimated CME propagation direction near the Sun with respect to the Earth, which is the average of the longitude and latitude from the GCS model. }
\tablenotetext{b}{Shock arrival time at the Earth. }
\tablenotetext{c}{Shock arrival time at \ulysses. }
\tablenotetext{d}{The latitude and longitude of \ulysses{} relative to the Earth and its distance from the Sun when the CME erupted. }
\tablenotetext{e}{The parameters of the 2001 November 22 CME are from \citet{2017Liu}. }
\tablenotetext{f}{The parameters of the 2005 May 13 CME are from \citet{2017Zhao}. }
\tablenotetext{g}{The parameters are from \citet{2008LiuL}, except the estimated CME propagation direction near the Sun. }
\end{table}

\clearpage

\begin{table}
\caption{Interplanetary propagation parameters of the 7 CMEs/shocks. \label{propa}}
\begin{tabularx}{\linewidth}{@{\extracolsep{\fill}}ccccccccc}
\hline\hline
Date   &${t}_{s}$\tablenotemark{a} & ${v}_{0}$\tablenotemark{b} &${v}_{s}$\tablenotemark{c} & ${v}_{s}$\tablenotemark{d}   & ${a}$\tablenotemark{e}  & ${t}_{a}$\tablenotemark{e} & ${r}_{a}$\tablenotemark{e} \\
   & (hr)  & (\kms)  & (\kms)  & (\kms)  & (m s$^{-2}$)  & (hr)  & (au) \\
\hline
\tablenotemark{1}1997 Nov 4                      & 64.4  & 1200  & 560   & 410   & ${-}$11.0  & 16.2  & 0.34 \\
\tablenotemark{2}2000 Jun 6                      & 41.4  & 1800  & 890   & 570   & ${-}$22.1  & 11.4  & 0.37 \\
\tablenotemark{3}2001 Apr 2                      & 41.0  & 2300  & 870   & 430   & ${-}$47.3  & 8.4   & 0.32 \\
\tablenotemark{4}2001 Nov 4                      & 33.5  & 2600  & 1040  & 920   & ${-}$52.6  & 8.2   & 0.36 \\
\tablenotemark{5}2001 Nov 22\tablenotemark{f}    & 30.9  & 2200  & 1050  & 900   & ${-}$20.8  & 15.4  & 0.60 \\
\tablenotemark{6}2005 May 13\tablenotemark{g}    & 33.6  & 2500  & 950   & 550   & ${-}$32.8  & 13.1  & 0.54 \\
\tablenotemark{7}2006 Dec 13\tablenotemark{h}    & 35.5  & 2200  & 1030  & 870   & ${-}$44.2  & 7.3   & 0.28 \\
\hline
\end{tabularx}
\tablenotetext{a}{Shock transit time from the Sun to the Earth. }
\tablenotetext{b}{CME/shock speed near the Sun estimated from a linear fit of the heliocentric distances from GCS. }
\tablenotetext{c}{Shock speed at the Earth. }
\tablenotetext{d}{Shock speed at \ulysses. }
\tablenotetext{e}{The deceleration, deceleration time period and deceleration cessation distance determined from the analytical model. }
\tablenotetext{f}{All parameters are from \citet{2017Liu}. }
\tablenotetext{g}{The shock speed near the Sun and its speeds at the Earth and \ulysses{} are provided by \citet{2017Zhao}. }
\tablenotetext{h}{The shock speeds at the Earth and \ulysses{} are taken from \citet{2008LiuL}. }
\end{table}

\clearpage

\clearpage

\clearpage

\clearpage

\clearpage

\clearpage

\clearpage

\clearpage

\clearpage

\clearpage

\clearpage

\clearpage

\clearpage

\clearpage

\clearpage

\clearpage

\clearpage

\clearpage

\clearpage

\clearpage

\clearpage

\clearpage

\clearpage

\bibliography{references}
\bibliographystyle{aasjournal}
\end{document}